\documentclass[sigconf,preprint,balance=false]{acmart}
\usepackage{popets}
\usepackage{subcaption}

\setcopyright{none}
\acmDOI{}
\acmISBN{}
\acmConference[]{}{}{}
\acmYear{}
\acmVolume{}
\acmNumber{}

\begin{document}

\title[Synth-MIA]{Synth-MIA: A Testbed for Auditing Privacy Leakage in Tabular Data Synthesis}


\author{Joshua Ward}
\affiliation{%
  \institution{University of California Los Angeles}
  \city{Los Angeles}
  \state{CA}
  \country{USA}}

\author{Xiaofeng Lin}
\affiliation{%
  \institution{University of California Los Angeles}
  \city{Los Angeles}
  \state{CA}
  \country{USA}}

\author{Chi-Hua Wang}
\affiliation{%
  \institution{University of California Los Angeles}
  \city{Los Angeles}
  \state{CA}
  \country{USA}}

\author{Guang Cheng}
\affiliation{%
  \institution{University of California Los Angeles}
  \city{Los Angeles}
  \state{CA}
  \country{USA}}


\renewcommand{\shortauthors}{Ward et al.}

\begin{abstract}
Tabular Generative Models are often argued to preserve privacy by creating synthetic datasets that resemble training data.  However, auditing their empirical privacy remains challenging, as commonly used similarity metrics fail to effectively characterize privacy risk. Membership Inference Attacks (MIAs) have recently emerged as a method for evaluating privacy leakage in synthetic data, but their practical effectiveness is limited. Numerous attacks exist across different threat models, each with distinct implementations targeting various sources of privacy leakage, making them difficult to apply consistently. Moreover, no single attack consistently outperforms the others, leading to a routine underestimation of privacy risk.

To address these issues, we propose a unified, model-agnostic threat framework that deploys a collection of attacks to estimate the maximum empirical privacy leakage in synthetic datasets. We introduce Synth-MIA \footnote{A
Repository is available at:\\ \href{https://github.com/joshward96/Synth-MIA}{https://github.com/joshward96/Synth-MIA}.\\}, an open-source Python library that streamlines this auditing process through a novel testbed that integrates seamlessly into existing synthetic data evaluation pipelines through a scikit-learn-like API. Our software implements 13 attack methods through a Scikit-Learn-like API, designed to enable fast systematic estimation of privacy leakage for practitioners as well as facilitate the development of new attacks and experiments for researchers.

We demonstrate our framework's utility in the largest tabular synthesis privacy benchmark to date, revealing that higher synthetic data quality corresponds to greater privacy leakage, that similarity-based privacy metrics show weak correlation with MIA results, and that the differentially private generator PATEGAN can fail to preserve privacy under such attacks. This underscores the necessity of MIA-based auditing when designing and deploying Tabular Generative Models. 
\end{abstract}


\maketitle

\section{Introduction}
Tabular data synthesis has seen success in numerous applications, including differentially private data release \cite{privbayes,yoon2018pategan,yoon2020anonymization}, training dataset augmentation for supervised learning \cite{cui2024tabulardataaugmentationmachine}, and missing value imputation \cite{zheng2023diffusionmodelsmissingvalue, liu2024self}. Recent advancements in deep learning have significantly improved the performance of synthetic data \cite{tabddpm,tabsyn,liu2023tabular} in these tasks. Alongside these improvements, increasingly sophisticated methods have been proposed to evaluate the quality of generated synthetic data, including Machine Learning Efficiency (MLE) tasks and statistical fidelity measures \cite{alaa2022faithful}. Yet, while methods to evaluate the \textit{success} of these models have evolved rapidly, approaches to diagnose model \textit{failure} have lagged behind.

Generative models are known to exhibit a tendency to overfit to regions of the training distribution, often memorizing specific training examples \cite{ganleaks,vanbreugel2023membership}. Indeed, previous studies \cite{Feldman2020, Feldman2020b, Brown_2021} have demonstrated that memorization is an intrinsic aspect of machine learning models and can be necessary to achieve optimal performance. This issue is exacerbated in the context of tabular data, which often contains sensitive or protected information about individuals in applications like Healthcare, Finance, and Education \cite{Giuffr2023HarnessingTP, Assefa, Flanagan22}. As tabular generative models grow larger and more expressive, there is a risk that their increasing success comes at the cost of privacy leakage from memorizing and overfitting to the training dataset \cite{vanbreugel2023membership,fang2025understanding}.

Recently, Membership Inference Attacks (MIAs) have seen growing interest in the machine learning community as their methodology specifically quantifies privacy risk as the result of a model failure \citep{Sablayrolles2019WhiteboxVB,Long,Carlini2021MembershipIA,watson2022on,ye2022,zarifzadeh2024lowcosthighpowermembershipinference}. MIAs pose privacy auditing as a game in which given limited information about a model and its outputs, an adversary attempts to classify if test data are members of the training dataset or are holdout data sampled from the same population. MIAs succeed in this game by exploiting the extent of model overfitting and memorization in a generated dataset and give an exact measure of individual privacy risk for each training observation \cite{Shokri,Carlini2021MembershipIA}.

However, deploying MIAs in practice has been difficult for the tabular data synthesis community. Implementations of synthetic data MIAs are scattered and can rely on specific model architectures and threat models, making them difficult to benchmark new methods with and deploy comparitive privacy audits. Different MIAs can also target different components of model failure (as we show in Table \ref{tab:mia_summary}) and so privacy leakage can be under-reported depending on what MIA is used. Indeed, there is not a strictly dominant MIA for tabular synthetic data (see Table \ref{tab:mia_performance} and Figure \ref{Attack Performance}) which implies that there is no guarantee an MIA successfully quantifies the maximum empirical privacy leakage of a synthetic dataset.

\textbf{Contributions}: To address these issues, we summarize our contributions as follows:
\begin{enumerate}
    \item We introduce a novel framework for MIA evaluation on tabular synthetic data where under a unified threat model, we deploy an ensemble of attacks to estimate maximum measures of privacy leakage. 
    \item We introduce Synth-MIA, an open-source Python library with a scikit-learn-like API designed to integrate this framework into existing synthetic data evaluation pipelines. Synth-MIA is a testbed that supports a comprehensive library of attacks, evaluation metrics, and features (see Table \ref{tab:comparison}) along with tutorials for privacy and tabular data synthesis-focused users alike to deploy audits and develop new attack methodologies.
    \item We demonstrate the usefulness of this framework and software in (to our knowledge) the largest tabular synthesis benchmark to date. We find that across 48 datasets and 9 architectures that there is a hidden cost to generating high quality synthetic data in that it corresponds to greater privacy leakage. We also find that commonly used non-adversarial privacy metrics are insufficient for understanding privacy leakage in synthetic data and show that MIA-based auditing can serve as a better replacement.
\end{enumerate}

\section{Preliminaries}

\subsection{Tabular Synthetic Data Generation}
We represent tabular data as a matrix $\mathbf{X} \in \mathcal{X}^{n \times d}$, where $n$ denotes the number of samples and $d$ the number of features. The set $\mathcal{X}$ defines the domain of possible feature values, which may span multiple data types. Each row $\mathbf{x}_i \in \mathcal{X}^d$ corresponds to a single observation, independently drawn from the true data distribution $p_X(X)$. The element $\mathbf{x}_{i,j}$ indicates the value of the $j$-th feature for the $i$-th observation. A training dataset $T = {\mathbf{x}_1, \mathbf{x}_2, \ldots, \mathbf{x}_n}$ is composed of these $n$ i.i.d. samples.

The objective of tabular generative modeling is to learn a model $G$ that captures the underlying distribution $p_X(X)$ using the training data $T$. Once trained, $G$ can produce synthetic samples $\tilde{\mathbf{x}} \sim G$, forming a synthetic dataset $S = {\tilde{\mathbf{x}}_1, \tilde{\mathbf{x}}_2, \ldots, \tilde{\mathbf{x}}_m}$. Ideally, the synthetic data generated by $G$ should closely reflect both the marginal distributions of individual features and the intricate joint dependencies among them, as found in the original dataset.

Tabular generative models offer the potential that their generated synthetic data can be private with respect to the original training examples. This privacy-preserving capability has driven significant research interest over the years, leading to diverse methodological approaches including Generative Adversarial Networks \cite{Xu2019ModelingTD, yoon2018pategan, yoon2020anonymization}, Variational Autoencoders \cite{Xu2019ModelingTD}, Normalizing Flows \cite{durkan2019neural}, large language Models \cite{ solatorio2023realtabformer} and diffusion models \cite{tabddpm, suh2023autodiffcombiningautoencoderdiffusion, tabsyn}.

\subsection{Membership Inference Attacks on Synthetic Data Generators}

Membership Inference Attacks (MIAs) represent a class of privacy attacks designed to determine whether a particular data point was included in the training dataset of a machine learning model. Consider a random variable $X$ defined over domain $\mathcal{X}$ following distribution $p_X(X)$, where $T$ represents a collection of independent samples drawn from this distribution. When a generative model $G$ is trained on dataset $T$ to produce synthetic dataset $S$, an adversary $\mathcal{A}: X \to \{0, 1\}$ seeks to identify whether a query sample $x^*$ belongs to the original training set $T$. 

This binary classification problem can be mathematically formulated as:
\begin{equation}\label{eq:membership_prediction}
\mathcal{A}(x^{\star}) = \mathbb{I}\left[f(x^{\star}) > \gamma\right]
\end{equation}
where $\mathbb{I}$ denotes the indicator function, $f(x^{\star})$ represents a scoring mechanism applied to the query sample $x^*$, and $\gamma$ serves as a decision boundary. The effectiveness of such attacks is typically evaluated using standard binary classification performance measures, providing insight into the degree of privacy breach concerning the training data.

The construction of these attacks depends critically on the adversary's knowledge and access level, commonly referred to as the threat model. No-box scenarios \cite{Hayes2017LOGANMI, Hilprecht2019MonteCA, ganleaks} assume access only to the synthetic dataset $S$. Calibrated attacks extend this by providing both $S$ and a reference dataset $R$ sampled from the same population distribution as the training data \cite{vanbreugel2023membership,ward2024dataplagiarismindexcharacterizing, Gen-LRA}. White-box attacks \citep{sablayrolles2019white} grant the adversary complete model access alongside both $S$ and $R$. Additional threat models consider scenarios where attackers possess knowledge of the generator's architecture and implementation details without access to trained parameters \cite{groundhog, houssiau2022tapas,Meeus_2024}.
\begin{figure*}[!htbp]
    \centering
    \begin{subfigure}[b]{0.48\textwidth}
        \centering
        \includegraphics[width=\textwidth]{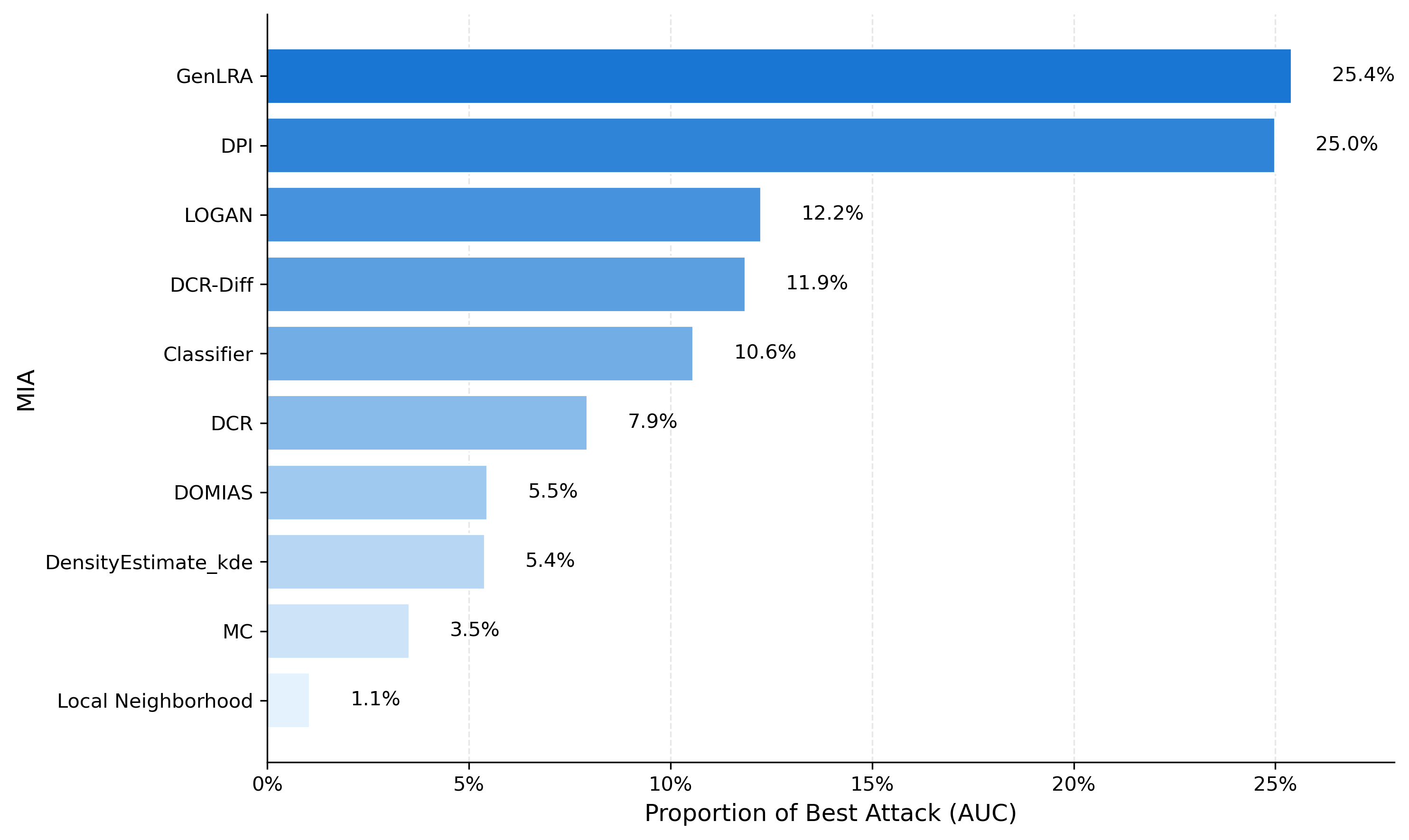}
        \label{fig:prop_best_auc}
    \end{subfigure}
    \hfill
    \begin{subfigure}[b]{0.48\textwidth}
        \centering
        \includegraphics[width=\textwidth]{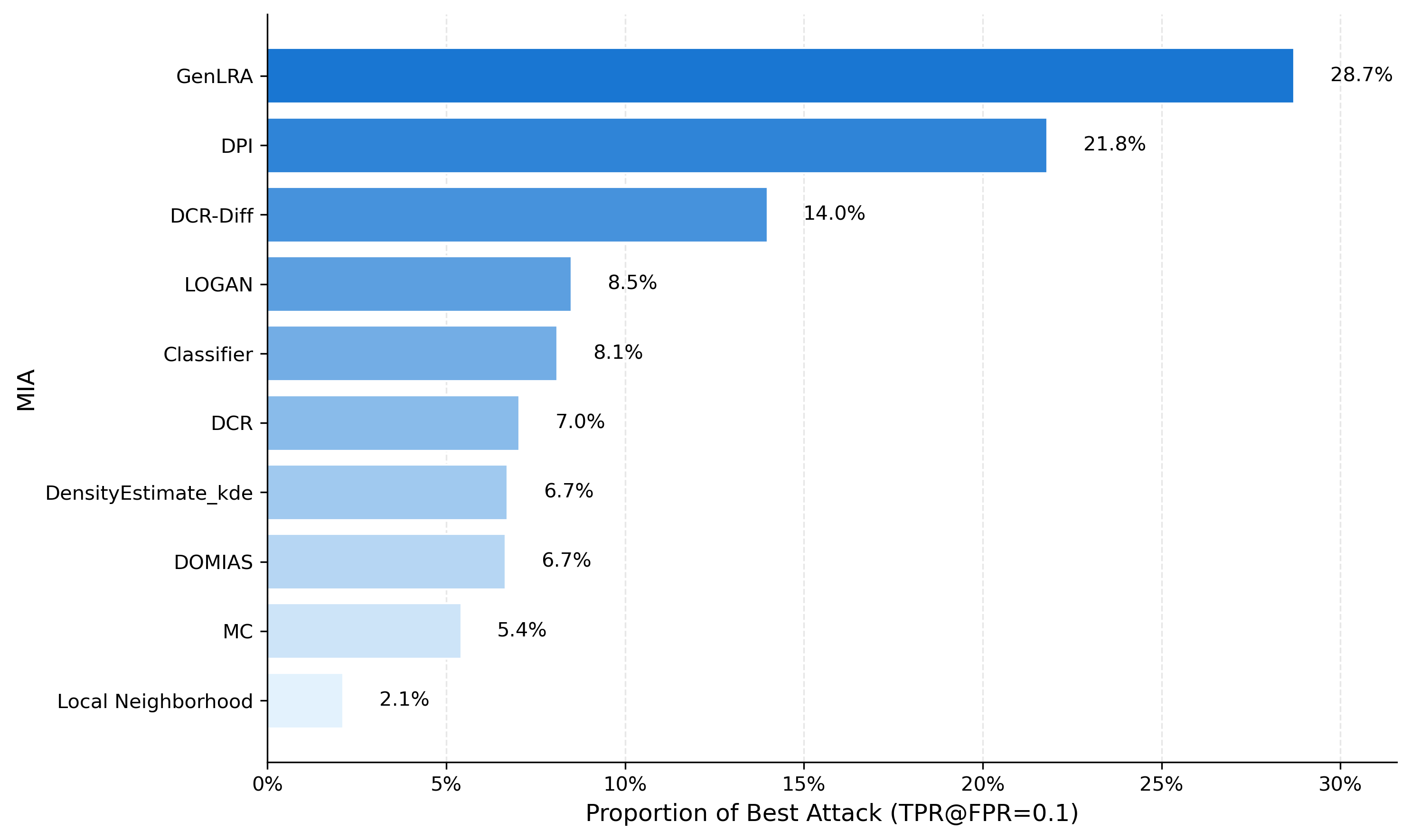}
        \label{fig:Attack Performance}
    \end{subfigure}
    \caption{Proportion of best attack across all generative models, datasets for (left) AUC scores and (right) True Positive Rate at False Positive Rate = 0.1. While Gen-LRA and DPI often perform the best, they are not strictly dominant- an auditing strategy that used just one method would routinely underestimate privacy.}
    \label{Attack Performance}
\end{figure*}
These attacks exploit different failure modes of generative models to construct the scoring function in Equation \ref{eq:membership_prediction}. One category targets memorization phenomena, where attacks from \cite{ganleaks} and \cite{houssiau2022tapas} calculate proximity metrics between the query point $x^*$ and its nearest neighbor in $S$. Another category focuses on distributional overfitting, where synthetic data exhibits excessive similarity to the training distribution compared to the broader population. Approaches including DOMIAS \cite{vanbreugel2023membership}, DPI \cite{ward2024dataplagiarismindexcharacterizing}, and Gen-LRA \cite{Gen-LRA} exploit this vulnerability by analyzing local density patterns in synthetic data relative to reference distributions. 

While methodologically diverse, MIAs targeting synthetic data aim to uncover the same fundamental issue: the potential for generative models to inadvertently reveal information about their training data. If a model produces synthetic records that allow an adversary to infer training membership, it constitutes a direct breach of privacy. 

 \subsection{Challenges with MIAs in Privacy Auditing Synthetic Data}
 MIAs are well-suited for auditing privacy leakage as they use interpretable classification metrics that highlight real material risk in the synthetic data generated by a model, as the success of the adversary implies the failure of the data. However, MIAs for tabular data synthesis suffer from two main challenges:

 \subsubsection{Use of different threat models}
Threat models significantly impact the performance and applicability of attacks on tabular data generators, yet there is no consensus in the literature on the appropriate threat model for synthetic data membership inference attacks. This lack of agreement poses a fundamental challenge: the interpretation and plausibility of privacy leakage depend on the assumed capabilities of the adversary, making comparisons between studies difficult and leading to potentially misleading conclusions about privacy risks.

For example, attacks such as \cite{groundhog, houssiau2022tapas, Meeus_2024} adopt a strong adversarial assumption in which the attacker has all knowledge of the generative model except training weights, including its architecture, training procedure, and potentially even hyperparameters. In contrast, works like \cite{ganleaks, vanbreugel2023membership, ward2024dataplagiarismindexcharacterizing, Gen-LRA} consider a more restricted attacker who lacks this information and must infer membership solely from synthetic data outputs and a reference dataset. This divergence in assumptions affects not only the success of each attack but also their practical implications: Synthetic data that appear highly unprivate under a "Model Known" Shadow-box threat model may be much stronger under a "Model Unknown" model, and vice versa.

\begin{figure*}
    \centering
    \includegraphics[width=1\linewidth]{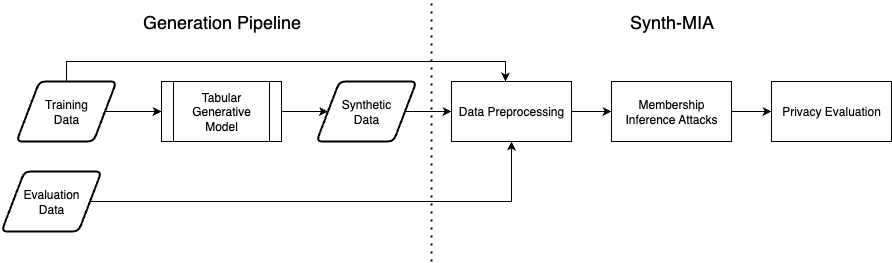}
    \caption{System diagram of Synth-MIA. An advantage of Synth-MIA is that it is model-agnostic and can be easily appended on to existing synthetic data pipelines. Users pass training, synthetic, and holdout datasets to a preprocessing step before initializing MIAs with a standard API. The evaluation step can then be called following all attacks.}
    \label{fig:system_diagram}
\end{figure*}

\subsubsection{Attack Disagreement}
An additional problem is that MIAs use different sources of model failure as signal and can thus disagree with leakage estimation, even using the same threat model. In Figure \ref{Attack Performance}, we demonstrate that the estimated privacy leakage from various MIAs differs and the best attack is not consistent across datasets. For a given MIA, there is no guarantee that it will perform well relative to other potential attacks as a model architecture and training scheme may exhibit more or less of a certain failure behavior an attack targets depending on the dataset. As no attack is strictly better across all possible situations, there is a problem for the defender in that auditing their generated synthetic data with only one method may underestimate the actual privacy leakage from an attacker that could use any method available.

\section{Synth-MIA: Auditing Framework} \label{sec:framework}
\subsection{Threat Model} 
In this work, we specifically focus on attacks designed under "No-box" and "No-box-calibrated" threat models \cite{houssiau2022tapas}.\footnote{Note: Depending on the source, the names for these threat models are used inconsistently across the literature. \cite{ganleaks, vanbreugel2023membership, ward2024dataplagiarismindexcharacterizing} for example refer to these as Black-box and Shadow-box or Calibrated Black-box threat models- always assuming the generator to be unknown. \cite{groundhog} on the other hand, uses these same names but considers the generator to be known; however, knowledge of the parameterization is restricted. \cite{houssiau2022tapas} instead distinguishes knowledge of the model as No-box vs Black-box.} 
 In these scenarios, we assume an adversary only has access to the synthetic data $S$ in the No-box case or an additional reference dataset $R$ sampled from the same population distribution as the training data in No-box-calibrated case. We assume the adversary has no knowledge of the generative model used to generate $S$ including its name, implementation, parameterization, etc. As a classic example of this threat model, an insurance company could have access to a hospital's public synthetic cancer dataset and, for a new applicant, attack the dataset to determine if the applicant is a member, leaking their diagnosis \citep{hu2022membershipinferenceattacksmachine}. In this scenario, the adversary would have no knowledge of the generative model the hospital used, but would have the synthetic data and perhaps a reference dataset from its internal records. We argue that these threat models should be the primary focus in the tabular data synthesis domain for the following reasons:

\textbf{Plausibility:} No-box and No-box-calibrated threat models best reflect realistic synthetic data release, where attackers only access the released data and possibly a reference dataset from domain knowledge, public sources, or purchased data. Even artificial references, such as histogram-based datasets, have been shown to improve attack performance \cite{vanbreugel2023membership}. By contrast, shadow-box “Model Known” attacks are less realistic: defenders can trivially defeat them by withholding model details, and disclosure has been shown to increase substantially increase leakage risk even in differentially private generators \cite{golob2024privacyvulnerabilitiesmarginalsbasedsynthetic}. Thus, best practice is to minimize model disclosure, making model-agnostic No-box and No-box-calibrated attacks the most proximal threat models to realistic security scenarios.

\textbf{Compatibility:} A key advantage of these threat models are that corresponding attacks are definitionally compatible with all tabular generators as they only assess the output of these models. This allows for fair benchmarking between both attacks and models and represents a data-centric approach similar to the corresponding utility metrics used for tabular data synthesis. 

\textbf{Cost:} Previous "Known Model" Shadow Box attacks are computationally expensive as they rely on training many surrogate models for each $x^*$ to construct their attack. In practice, these approaches are impractical as they require training many separate models to audit a single trained model, especially as large diffusion and language model architectures become more popular. In contrast, No-box and No-box-calibrated attacks are significantly more cost-effective. For example, the most expensive attack in this category, Gen-LRA, requires fitting only separate density estimators, which is far less computationally intensive. This efficiency makes No-box and No-box-calibrated attacks more practical for large-scale privacy audits and real-world applications.

\subsection{Auditing Procedure}
Under the proposed threat model, the auditor faces a methodological challenge: no single attack strictly dominates all others in terms of effectiveness. If such an attack existed, the synthetic data could be audited using this attack to establish an empirical privacy bound. In the absence of an optimal attack, we propose a comprehensive audit by simply evaluating a broad set of membership inference attacks (MIAs) and reporting the maximum worst-case measure of privacy leakage.

This approach aligns with the principles of Empirical Differential Privacy (EDP) \cite{Jagielski2020}, wherein the privacy loss of a synthetic dataset is characterized by the supremum of privacy leakage across all possible MIAs. Specifically, the empirical privacy level of a generative model is determined by the most effective attack strategy. Formally, let $\mathcal{A} = \{A_1, A_2, \dots, A_n\}$ denote a set of MIAs. The effective epsilon level is given by:

$$
\sup_{A \in \mathcal{A}} \text{TPR}_A(\tau) \quad \text{subject to } \text{FPR}_A(\tau) \leq \delta.
$$

where $\tau$ is some decision threshold. This means that the worst-case privacy leakage is estimated by selecting the most effective MIA, which naturally leads to reporting the maximum observed attack performance across different strategies. Since different MIAs exploit different aspects of overfitting, such as direct memorization, attribute leakage, and distributional biases, this approach ensures that privacy assessments are robust.
\begin{table*}[t]
\small
\centering
\caption{A taxonomy of No-box and No-box-calibrated MIAs for synthetic data. All attacks are available within the Synth-MIA library. The scoring function is the defined $f(x^*)$ from Equation \ref{eq:membership_prediction} from each attack.}
\begin{tabular}{@{}lllll@{}}
\toprule
\textbf{MIA} & \textbf{Source} & \textbf{Signal Motivation} & \textbf{Threat Model} & \textbf{Scoring Function} \\
\midrule
Classifier & \cite{houssiau2022tapas} & Overfitting & Calibrated & $p_S(x^*) / p_R(x^*)$ \\
Density Estimator & \cite{houssiau2022tapas} & Memorization & No-box & $p_S(x^*)$ \\
DCR & \cite{ganleaks} & Memorization & No-box & $-\min_{{x} \in S} d(x^*,{x})$ \\
DCR-Diff & \cite{ganleaks} & Memorization & Calibrated & $-\min_{{x} \in S} d(x^*,{x}) - \min_{{x} \in R} d(x^*,{x})$ \\
DOMIAS-BNAF & \cite{vanbreugel2023membership} & Overfitting & Calibrated & $p_S(x^*) / p_R(x^*)$ \\
DOMIAS-KDE & \cite{vanbreugel2023membership} & Overfitting & Calibrated & $p_S(x^*) / p_R(x^*)$ \\
DPI & \cite{ward2024dataplagiarismindexcharacterizing} & Memorization & Calibrated & $\frac{\sum_{\mathbf{z} \in D(x^*)} \mathbb{I}(\mathbf{z} \in S)}{\sum_{\mathbf{z} \in D(x^*)}\mathbb{I}(\mathbf{z} \in R)}$ \\
Gen-LRA-BNAF & \cite{Gen-LRA} & Overfitting & Calibrated & $p_{R \cup x^*}(S) / p_R(S)$ \\
Gen-LRA-KDE & \cite{Gen-LRA} & Overfitting & Calibrated & $p_{R \cup x^*}(S) / p_R(S)$ \\
Local Neighborhood & \cite{houssiau2022tapas} & Memorization & No-box & $knn_S(x^*)$ \\
Logan & \cite{Hayes2017LOGANMI} & Overfitting & Calibrated & $p_S(x^*) / p_R(x^*)$ \\
MC Estimation & \cite{Hilprecht2019MonteCA} & Memorization & No-box & $p_S(x^*)$ \\
\bottomrule
\end{tabular}

\label{tab:mia_summary}
\end{table*}
Furthermore, if the defender makes the strong assumption that the attacker is limited to selecting from the set of attacks known to the defender, this search process definitionally establishes an exact bound on the maximal privacy leakage achievable to an attacker if the synthetic data were to be released. This assumption allows the defender to systematically evaluate the worst-case privacy risk associated with their dataset.

\section{Synth-MIA: Software}
We introduce Synth-MIA, an open-source software framework designed to facilitate comprehensive privacy auditing in tabular data synthesis. Synth-MIA provides a robust and accessible platform for deploying generator-agnostic Membership Inference Attacks (MIAs) within a Python API modeled after Scikit-Learn. The framework currently supports 13 attack methods, incorporates all standard MIA evaluation metrics, and offers a suite of utility functionality that enables  integration into existing synthetic data assessment pipelines.

The framework enables efficient deployment of multiple MIA methods against synthetic datasets, systematic analysis and quantification of privacy risks, and a flexible environment for testing novel attacks. Operating independently of model training and sample generation, Synth-MIA is designed to be lightweight and customizable: it accepts training, holdout, and synthetic, and reference datasets as inputs, preprocesses them, executes MIA methods, and evaluates privacy risks. The software supports multiple deployment options, can easily integrate into existing research and industry pipelines, and includes tutorials to support benchmarking generative models, developing new attacks, and conducting comprehensive privacy audits.

\subsection{System Design}
Synth-MIA follows a modular system design organized around three core components: data preprocessing, attack execution, and privacy evaluation. This separation simplifies user interaction and ensures that attacks can be deployed, extended, and compared in a safe, replicable manner.

\textbf{Preprocessing module.} Attacks in Synth-MIA operate on training, holdout, synthetic, and reference datasets. To standardize these inputs, the preprocessing layer handles type casting, normalization, and feature encoding across all datasets. Importantly, encoders are fit to either the synthetic or reference datasets, preventing information leakage that would give the adversary an unrealistic advantage. This design makes preprocessing both safer and more convenient, allowing users to prepare datasets without risking biased or invalid results.
\begin{table*}[t]
\small  
\centering
\caption{A Comparison of Synth MIA, Tapas, and Synthcity. Synth-MIA is designed to be both a package that can be easily integrated into existing synthetic data evaluation pipelines and a tool for privacy researchers to develop new attack methods with. Synth-MIA thus has a wide array of features implemented to make it flexible to specific use-cases as well as the largest collection of attacks to support benchmarking and privacy auditing.}
\begin{tabular}{llccc}
\toprule
\textbf{Category} & \textbf{Feature} & \textbf{Synth MIA} & \textbf{Tapas} & \textbf{Synthcity} \\
\midrule
{MIAs} & DCR & \checkmark & \checkmark &  \\
 & Density Estimator & \checkmark & \checkmark &  \\
 & Local Neighborhood & \checkmark & \checkmark &  \\
 & DCR-Diff & \checkmark &  &  \\
 & DOMIAS-KDE & \checkmark &  & \checkmark \\
 & DOMIAS-BNAF & \checkmark &  & \checkmark \\
 & DPI & \checkmark &  &  \\
 & LOGAN & \checkmark &  &  \\
 & MC & \checkmark &  &  \\
 & Gen-LRA-KDE & \checkmark &  &  \\
 & Gen-LRA-BNAF & \checkmark &  &  \\
 & Classifier & \checkmark &  &  \\
 & Groundhog &  & \checkmark &  \\
\midrule
{Features} & Tutorial Notebooks & \checkmark &  & \checkmark \\
 & Custom Attack Developer Templates & \checkmark &  &  \\
 & Attack Formulation Helper Functions & \checkmark &  &  \\
 & Flexible Modular Pre-processing & \checkmark &  &  \\ 
 & Generator Agnostic & \checkmark &  &  \\
  & Customizable metrics/report/mixture of attack & \checkmark &  &  \\
 & PIP Deployment & \checkmark &  & \checkmark \\
 & Docker Deployment & \checkmark &  &  \\
 & Support of large datasets via Spark/Dask & \checkmark &  &  \\
\midrule
{Metrics} & AUC & \checkmark & \checkmark & \checkmark \\
 & Accuracy & \checkmark & \checkmark &  \\
 & TPR & \checkmark & \checkmark &  \\
 & FPR & \checkmark & \checkmark &  \\
 & Advantage & \checkmark & \checkmark &  \\
 & Privacy Gain & \checkmark & \checkmark &  \\
 & Effective Epsilon & \checkmark & \checkmark &  \\
 & TPR@FPR & \checkmark &  &  \\
 & Brier Score & \checkmark &  &  \\
\bottomrule
\end{tabular}

\label{tab:comparison}
\end{table*}

\textbf{Attack module.} All attack implementations inherit from a common \texttt{BaseAttacker} class, modeled after the Scikit-Learn estimator interface. This abstraction eliminates boilerplate, enforces a uniform workflow, and enables direct comparison of methods under consistent conditions. Attacks are initialized with hyperparameters, executed via \texttt{MIA.attack()} on preprocessed datasets, and return predicted scores along with ground-truth labels for training and holdout samples. New attacks can be introduced simply by subclassing, without re-implementing shared boilerplate. 

\textbf{Evaluation module.} After attacks are executed, Synth-MIA provides an evaluation module to quantify privacy risk. The suite implements ROC-based metrics such as AUC-ROC and TPR@FPR, thresholded classification metrics including accuracy, precision, and recall, and also supports estimation of the empirical privacy parameter $\epsilon$ following \cite{Jagielski2020}. Results are returned as standardized Python dictionaries, enabling straightforward comparison of privacy risks across datasets and attack methods.

Together, these modules form a cohesive framework that is both flexible and reliable. As membership inference auditing requires accuracy and reproducibility, Synth-MIA is implemented with attention to efficiency, safety, and software rigor. Core routines follow standard design patterns (e.g., estimator-style interfaces, modular abstractions), making components reusable and extensible while reducing implementation errors. The codebase incorporates typing, unit and integration tests, and careful preprocessing with schema validation to guard against information leakage or unsafe transformations. These practices ensure that Synth-MIA is not only easy to use and extend, but also provides trustworthy and reproducible privacy risk estimates.

\subsection{Included Attacks}
Synth-MIA supports a comprehensive list of every black and shadow box attack that has to our knowledge been proposed for tabular synthetic data. These cover a wide array of attack strategies focused on evaluating the synthetic data for overfitting and memorization. We present a taxonomy of these attacks in Table \ref{tab:mia_summary} where each is implemented in our library and further briefly summarize them.

\textbf{Distance to Closest Record (DCR / DCR-Diff). }Distance-based membership inference attacks \cite{ganleaks} are based on the intuition that synthetic data models may memorize training examples, leading to synthetic samples that lie closer in feature space to training members than to non-members. The Distance to Closest Record (DCR) method \cite{ganleaks} formalizes this intuition by defining
$f_{\text{DCR}}(x^*,S) = -\min_{{x} \in S} d(x^*,{x})$,
where $d(\cdot,\cdot)$ is a chosen distance metric. The variant DCR-Diff extends this by incorporating a reference dataset $R$, adjusting the score by subtracting the distance to the nearest reference record:
$f_{\text{DCR}}(x^*,S,R) = -\min_{{x} \in S} d(x^*,{x}) - \min_{{x} \in R} d(x^*,{x})$.

\textbf{DOMIAS / Density Estimation.} The DOMIAS attack \cite{vanbreugel2023membership} takes a density-based perspective, detecting overfitting signals within synthetic datasets. It computes a density ratio between synthetic and reference distributions, yielding the scoring function
$f_{\text{DOMIAS}}(x^*,S,R) = \frac{p_S(x^*)}{p_R(x^*)}$.
This requires separate density estimation of $S$ and $R$, which can be achieved via kernel density estimation or neural density estimators. \cite{houssiau2022tapas} a presents a simpler strategy of estimating the density of $x^*$ over the synthetic dataset: $f_{\text{Density Estimate}}(x^*,S) = {p_S(x^*)}$

\textbf{Data Plagiarism Index (DPI) / Local Neighborhood.} The Data Plagiarism Index \cite{ward2024dataplagiarismindexcharacterizing} measures local memorization effects by examining neighborhood density ratios. For $x^*$, DPI forms a $k$-nearest-neighbor set $D(x^*)$ containing both reference and synthetic points, then computes
$f_{\text{DPI}}(x^*, S, R) = \frac{\sum_{\mathbf{z} \in D(x^*)} \mathbb{I}(\mathbf{z} \in S)}{\sum_{\mathbf{z} \in D(x^*)}\mathbb{I}(\mathbf{z} \in R)}$.
A related approach, Local Neighborhood \cite{houssiau2022tapas}, instead counts neighbors within a radius around $x^*$.

\textbf{Gen-LRA.} Gen-LRA \cite{Gen-LRA} frames membership inference in terms of the influence of a query record $x^*$ on the likelihood of $S$, as estimated by a surrogate density model trained on $R$. If including $x^*$ substantially increases the likelihood of $S$, this is taken as evidence of memorization. The method further restricts evaluation to samples near $x^*$ and applies Gaussian KDE for density estimation. The resulting score is
$f_{\text{Gen-LRA}}(x^*,S,R) = \frac{\prod_{s \in S} p_{R \cup {x^*}}(s)}{\prod_{s \in S} p_R(s)}$. Gen-LRA further uses a K-Nearest-Neighbor based filter to select which synthetic samples it evaluates over for each $x^*$ where $k$ is a hyperparameter.

\textbf{LOGAN / Classifier-based.} Originally proposed as a white-box attack \cite{Hayes2017LOGANMI}, LOGAN was later adapted into a black-box form in \cite{vanbreugel2023membership}. The approach trains a GAN on synthetic data, using its discriminator $D_\theta(x)$ to separate target-generated samples $S$ from reference samples $R$. The discriminator score $f_{\text{LOGAN}}(x^*,S,R) = D_\theta(x^*)$ serves as the attack function, under the premise that members are more likely to be classified as synthetic. A refinement by \cite{houssiau2022tapas} replaces the GAN discriminator with a supervised classifier, e.g., Random Forests.

\begin{table*}[h]
\caption{Mean privacy metrics associated with the best performing MIA for each model architecture over all datasets and seeds. For each metric, lower is better and we bold the highest values. RealTabFormer (RTF) which is a state-of-the-art tabular generator produces synthetic data with on average the greatest privacy risk. The TPR@FPR=0 values also show that each architecture exhibits behavior that makes ~2-5\% of the training set perfectly detectable (see \cite{Carlini2021MembershipIA} for a thorough review of TPR@FPR).}

\begin{tabular}{lcccccc}
\toprule
& \multicolumn{5}{c}{\textbf{Max-MIA Performance Metrics}} \\
\cmidrule(lr){2-6}
 \textbf{Model} & \textbf{AUC-ROC} & \textbf{TPR@FPR=0} & \textbf{TPR@FPR=0.001} & \textbf{TPR@FPR=0.01} & \textbf{TPR@FPR=0.1} & \textbf{DCR-Prop} \\
\midrule
AdsGAN & 0.546 (0.035) & 0.029 (0.051) & 0.030 (0.051) & 0.041 (0.048) & 0.154 (0.041)  & 0.751 (0.16) \\
ARF & 0.564 (0.041) & 0.033 (0.063) & 0.034 (0.063) & 0.048 (0.061) & 0.175 (0.055)  & \underline{0.765 (0.153)} \\
CT-GAN & 0.544 (0.032) & 0.028 (0.049) & 0.028 (0.049) & 0.039 (0.047) & 0.154 (0.043)  & 0.752 (0.158) \\
Tab-DDPM & 0.548 (0.042) & 0.027 (0.049) & 0.028 (0.048) & 0.039 (0.046) & 0.156 (0.047)  & 0.752 (0.164) \\
N-Flow & 0.538 (0.027) & 0.026 (0.046) & 0.027 (0.046) & 0.037 (0.044) & 0.146 (0.035)  & 0.754 (0.151)\\
PATEGAN ($\epsilon =1$) & 0.536 (0.035) & 0.027 (0.063) & 0.028 (0.063) & 0.038 (0.061) & 0.148 (0.056)  & 0.742 (0.192)\\
RTF & \textbf{0.594 (0.060)} & \textbf{0.053 (0.088)} & \textbf{0.054 (0.088)} & \textbf{0.072 (0.085)} & \textbf{0.224 (0.098)}  & \textbf{0.774 (0.158)}\\
TabSyn & \underline{0.569 (0.057)} & \underline{0.043 (0.089)} & \underline{0.044 (0.089)} & \underline{0.058 (0.086)} & \underline{0.186 (0.086)}  & 0.758 (0.166) \\
TVAE & 0.551 (0.032) & 0.030 (0.052) & 0.030 (0.052) & 0.044 (0.050) & 0.166 (0.046)  & 0.758 (0.161)\\
\bottomrule
\end{tabular}
\label{tab:metrics_privacy}
\end{table*}

\textbf{Monte Carlo (MC).} The Monte Carlo attack \cite{Hilprecht2019MonteCA} probes overfitting by counting how often synthetic samples fall near a query. Defining the $\varepsilon$-neighborhood around $x^*$ as $U_\varepsilon(x^*) = {x' \mid d(x^*, x') \leq \varepsilon}$, the method estimates the probability mass in this region by drawing $n$ samples $s_1,\ldots,s_n$ from $S$ and computing
$f_{\text{MC}}(x^*,S) = \frac{1}{n} \sum_{i=1}^{n} \mathbb{I}(s_i \in U_\varepsilon(x^*))$.

\subsection{Comparison to Other Software}
A variety of other software packages have been developed for privacy auditing tabular generative models and synthetic data. These include TAPAS \cite{houssiau2022tapas} which directly focuses on MIAs, Anonymeter \cite{anonymeter} which audits privacy using a linkage and attribute inference framework, and then a variety of other packages such as Synthcity are more focused on generating synthetic data that then feature some MIAs in the evaluation methodology \cite{synthcity, du2024systematicassessmenttabulardata}. Of these, Anonymeter is not focused on MIA-based auditing and other packages developed for the entire synthetic data pipeline are too general for specific auditing and benchmarking purposes in that they are often missing important features and metrics (See table \ref{tab:comparison} for a full comparison of featured attacks, features, and metrics).

The most proximate package to Synth-MIA is TAPAS which is dedicated to deploying Membership Inference Attacks against tabular generative models. The main difference between Synth-MIA and TAPAS is that Synth-MIA is focused on the model agnostic threat model setting whereas TAPAS explicitly requires that the model architecture and training parameters are passed to it. From a privacy perspective, this threat model is not realistic for the reasons we highlight in Section \ref{sec:framework}, and additionally, this requirement can create friction for users who need to connect any new model architecture to this package in order to perform privacy auditing. As Table \ref{tab:comparison} shows, Synth-MIA has a greater diversity of attacks, features, and metrics.

\section{Experiments} 
\label{sec:experiments}

A key use-case of Synth-MIA is benchmarking the privacy of tabular generative models, as existing benchmarks often lack comprehensive membership inference attack evaluations \cite{tabddpm,tabsyn,solatorio2023realtabformer}. In this section, we evaluate the effectiveness of the attacks summarized in Table \ref{tab:mia_summary} and the privacy audit scheme of Section \ref{sec:framework} on the synthetic data of common tabular generative models. Specifically, we investigate how MIA success relates to standard measures of data utility, fidelity, and privacy, providing insight into the trade-offs between these properties. In addition, we explore whether these models exhibit common patterns of privacy leakage, shedding light on their vulnerabilities in real-world deployment scenarios.

\subsection{Baselines}
We evaluate the effectiveness of Synth-MIA in what we believe to be the largest synthetic data privacy benchmark to date. Our primary objective is to assess the privacy performance of widely used tabular generative models across a diverse range of datasets.

\textbf{Chosen Models.} The tabular generative models included in our evaluation span a wide array of architectures. CT-GAN, TVAE \cite{Xu2019ModelingTD}, Normalizing Flows (N-Flows) \cite{durkan2019neural}, Adversarial Random Forests (ARF) \cite{pmlr-v206-watson23a}, Tab-DDPM \cite{tabddpm}, AdsGAN \cite{yoon2020anonymization} are all models frequently used in synthetic data benchmarking and we use the default hyperparameters from their implementations in Synthcity \cite{synthcity}. We also include the more recent, state of the art models TabSyn \cite{tabsyn} and Realtabformer (RTF) \cite{solatorio2023realtabformer} of which we use their default hyperparameters from their native implementations. Lastly, we include the differentially private data generator PATEGAN \cite{yoon2018pategan} to study how differential privacy can offer protection against these attacks. We initialize the Synthcity implementation for PATEGAN using the default hyperparameters but set three levels of $\epsilon$ at 1, 5, and 10.

\textbf{Datasets.} Our benchmark spans 48 datasets taken from the OpenML-CC18 Curated Classification benchmark \cite{oml-benchmarking-suites} representing a broad variety of fields including economics, healthcare, and social sciences. As the original benchmark contains 72 datasets, we filter out instances that have greater than 100 columns as not all models can successfully handle such high dimensionality. Following standard synthetic data benchmarking practices \cite{synthcity, tabsyn}, the data is then split into 80:20 train/test partitions, and synthetic data is generated to match the training set size. We repeat this across 5 seeds where we follow the recommendation of \cite{guépin2024lostaveragesnewspecific}, fixing the train/test partition across all runs and varying only the generative model initialization seeds in order to prevent confounding in attack performance from differing data.

\textbf{Privacy Assessment.} We apply the Synth-MIA privacy auditing framework, estimating Max-AUC and Max-TPR@FPR across multiple Membership Inference Attacks. Here, we preprocess each dataset by scaling continuous variables and one-hot encoding categorical variables based on the synthetic data available to the adversary. For kernel density estimator based attacks, we ordinal-encode categorical variables to help with estimator convergence. For all attacks, we further split the test set into equal-sized holdout and reference datasets. Here, following \cite{vanbreugel2023membership,ward2024dataplagiarismindexcharacterizing} all training data are included for evaluation as the positive class with each MIA and all holdout data are evaluated and treated as the negative class. We use a maximum of 1000 training and 1000 holdout examples for the MIA evaluation set. We include a list of attacks and hyperparameters used in the Appendix for finding these maximum values.

\textbf{Synthetic Data Assessment.} To assess the quality of the generated data, we compute standard utility metrics, including classifier AUC from an XGBoost model \cite{xgboost} trained on the synthetic data and evaluated on the real holdout set. We also measure fidelity using Maximum Mean Discrepancy and Jensen-Shannon Distance, alongside the widely used privacy metric, Distance to Closest Record (DCR) proportion \cite{park2018data}.

This approach differs from prior work in the synthetic data MIA literature, which typically evaluates individual attacks across varying train/test splits and training dataset sizes \cite{vanbreugel2023membership,ward2024dataplagiarismindexcharacterizing,Gen-LRA}. While such methodologies are valuable for assessing the efficacy of new attacks, to our knowledge, no previous studies have systematically benchmarked MIAs for tabular generative models within the experimental framework commonly used in the tabular generation literature. In practice, synthetic data practitioners are unlikely to withhold half or two-thirds of their dataset solely for privacy evaluation, as is often implied in prior MIA studies. A key motivation for our experiment is to demonstrate the feasibility of using MIAs as a privacy auditing tool within the experimental constraints typical of tabular generative models— where member/non-member class distributions are naturally imbalanced, and reference datasets are not maximally large.

\subsection{Results}
As the utility and fidelity of the synthetic data generated from these models have been extensively studied \cite{synthcity,tabsyn}, we specifically focus our experimentation on understanding these models' relationships to privacy leakage estimated using Synth-MIA. Here, we take the maximum AUC and TPR@FPR values across all attacks for each generated synthetic dataset and aggregate the means and standard deviations across all seeds. This approach allows us to explore the behavior of tabular generative models, and how privacy leakage correlates with different aspects of synthetic data generation.

\begin{figure}
    \centering
    \includegraphics[width=1\linewidth]{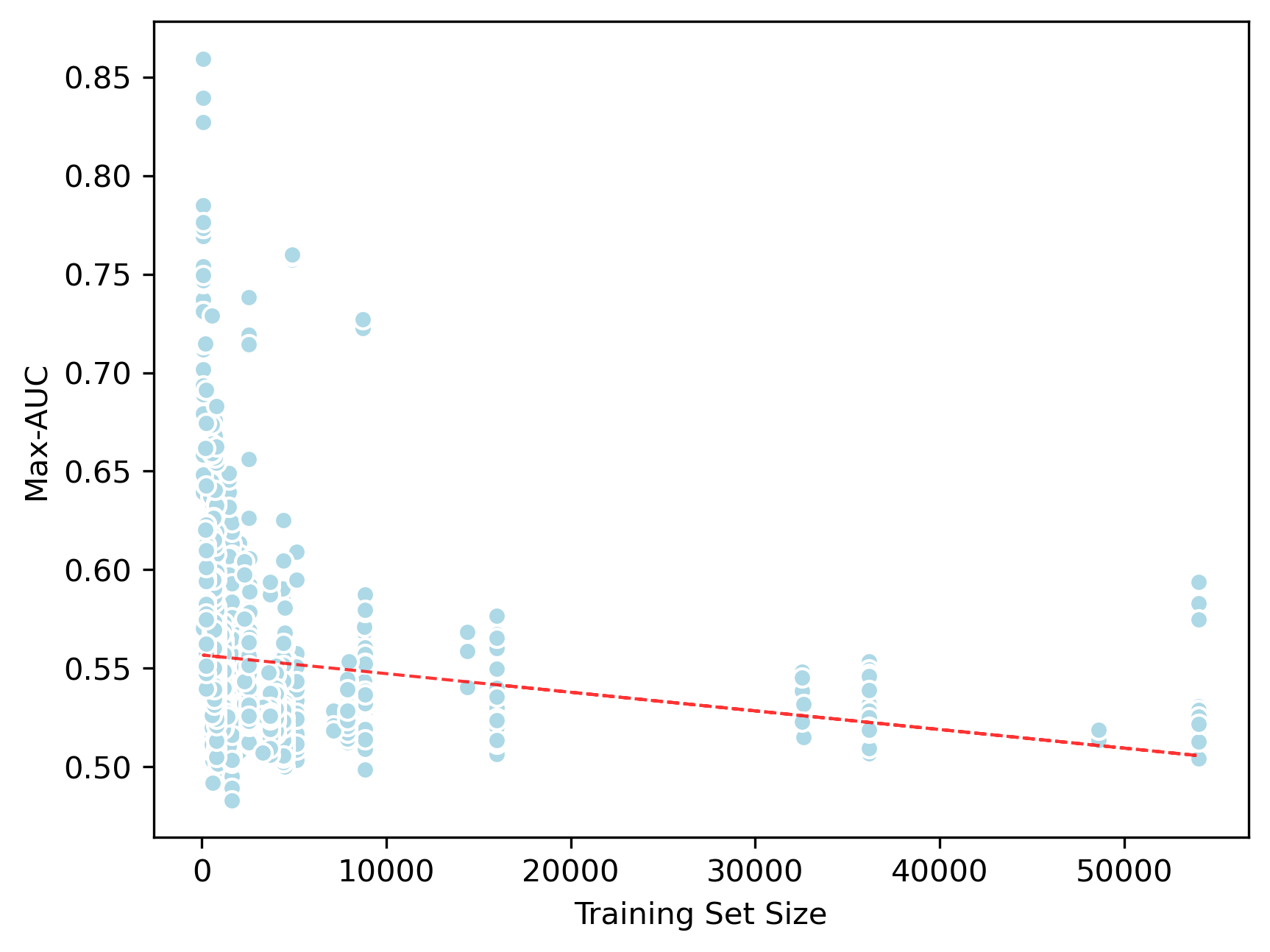}
    \caption{Max-AUC of all MIAs by training set N-size for all synthetic datasets. This plot suggests that synthetic data trained with fewer training observations exhibit a tendency to leak more privacy.}
    \label{fig:training_size}
\end{figure}
\begin{figure}
    \centering
    \includegraphics[width=1.0\linewidth]{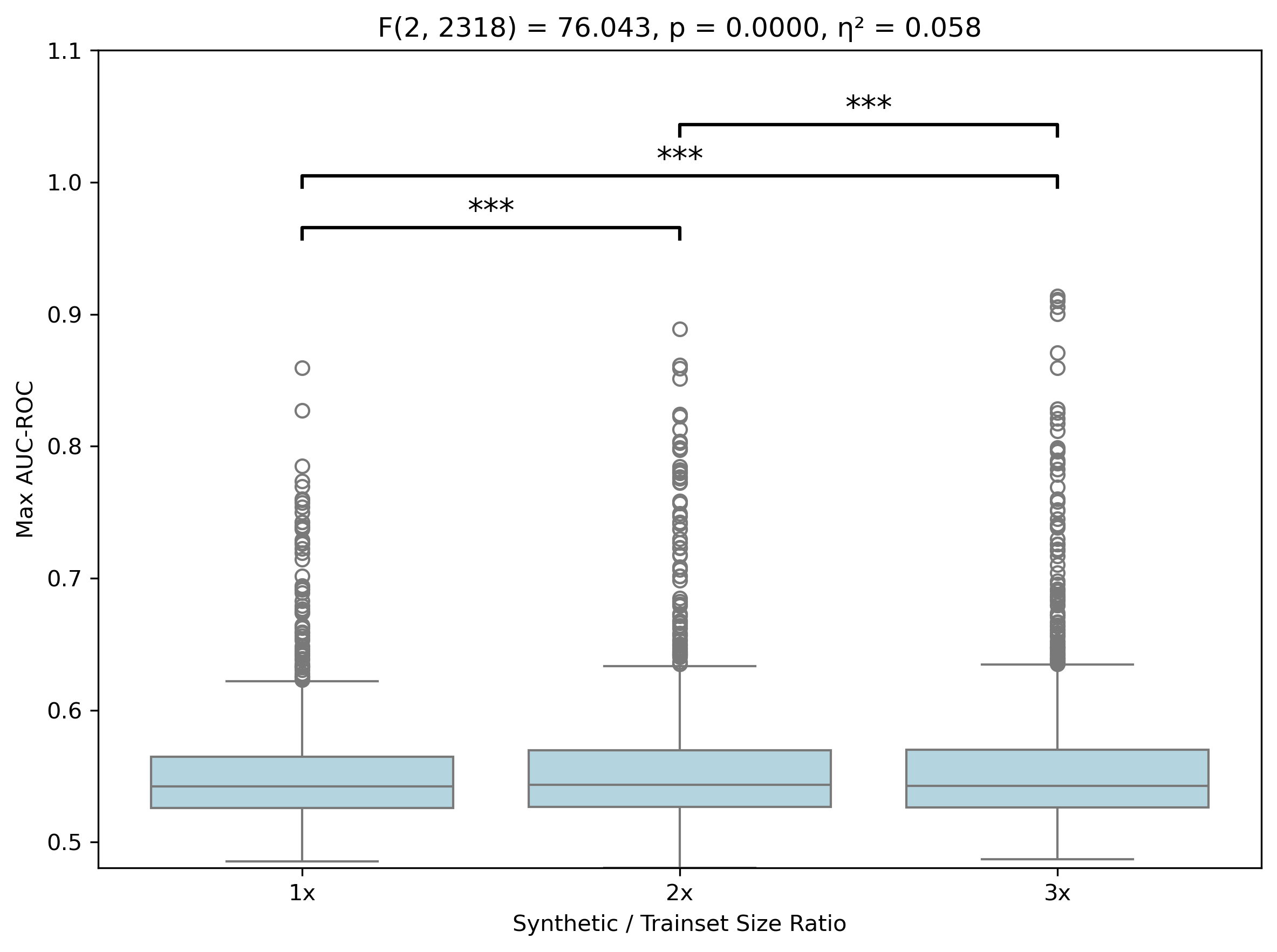}
    \caption{One-Way Repeated Measures ANOVA of Max-AUC for various synthetic data sizes. Per our experiment design, the synthetic dataset size is directly proportional to the size of the training dataset. For each run, we therefore generate synthetic dataset sizes that are 1, 2, and 3 times the size of the training set.}
    \label{fig:Anova}
\end{figure}

\subsubsection{Relationship of Privacy Leakage and Synthetic Data Quality}
\begin{figure*}
    \centering
    \includegraphics[width=1\linewidth]{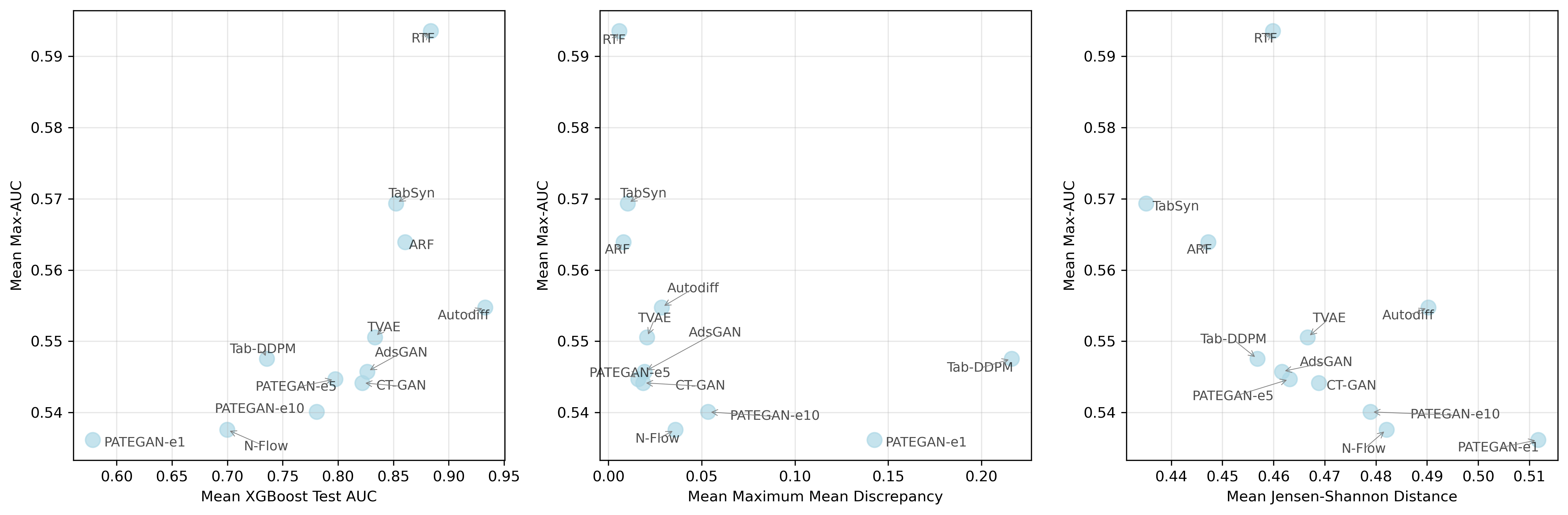}
    \caption{Privacy and quality trade-off across tabular generative model architectures. We plot the mean Max-AUC of all attacks for all datasets for each model architecture against corresponding measures of mean Classifier AUC, mean Maximum Mean Discrepancy, and mean Jensen-Shannon Distance. We find that models that produce synthetic data of higher quality also exhibit greater privacy leakage behavior. 
}
    \label{fig:privacy_utility}
\end{figure*}

We characterize a privacy leakage/ synthetic data quality trade-off among model architectures by comparing the mean Max-AUC for each model with corresponding utility (XGboost Test AUC) and fidelity metrics (Maximum Mean Discrepancy and Jensen-Shannon Distance) in Figure \ref{fig:privacy_utility}. As MIAs target memorization and overfitting to the training dataset, there is an implication that there is an implicit and often unreported privacy \textit{cost} to high fidelity and utility synthetic data. We also find that newer, larger models such as TabSyn and Realtabformer which are state of the art architectures usually produce the highest quality synthetic data while also exhibiting the highest Max-AUC and Max-TPR@FPR at all FPR levels. 
\subsubsection{Training Size and Privacy Leakage}
We are interested in finding common characteristics of the training datasets themselves that make it difficult for tabular data generators to produce leakage-free data. We compare the training sample size to the Max-AUC across all synthetic datasets in Figure~\ref{fig:training_size}. We find that synthetic datasets trained with smaller sample sizes tend to exhibit greater privacy leakage and that training set size is negatively correlated with Max-AUC ($r = -0.2390$, $p = 1.29 \times 10^{-24}$).
\subsubsection{Synthetic Size and Privacy Leakage}
We further investigate the effect of the generated synthetic set size on privacy leakage. Per our experiment design, the synthetic dataset size is directly proportional to the size of the training dataset. For each run, we therefore generate synthetic dataset sizes that are 1, 2, and 3 times the size of the original training dataset allowing us to analyze the effect of synthetic data size within each experimental unit. We thus conduct a One-Way Repeated Measures ANOVA between Max-AUC and the three levels of synthetic dataset size in Figure~\ref{fig:Anova}. We find that there are statistically significant differences between synthetic dataset level across post-hoc tests. However, the difference in means between each level is small and and $\eta^2$ of 0.058 suggests that the effect size is low.

\subsubsection{Attack Performance}
For each model, dataset, and seed we analyze which attack performed best from an AUC perspective. We plot the results in Figure \ref{Attack Performance} where we find that Gen-LRA is the best attack in 25.4\% and 28.7\% of the synthetic datasets for AUC-ROC and TPR@FPR=0.1 respectively. Indeed, we find that each attack outperformed the rest in at least some of the instances. 

In order to better understand the performance of individual MIAs relative to each other, we report the mean ranks over each experiment run for each attack in Table \ref{tab:mia_performance}. Here, we find that DPI and Gen-LRA compete between having the lowest mean ranks for AUC and various levels of TPR@FPR. However, the standard deviations of the ranks for each attack at each metric indicate that there is substantial variability in attack performance across models, datasets, and seeds.

This highlights a motivation of this paper in that since no attack dominates the others, the best course of action is to audit using all available attacks. We also find that the top three attacks of Gen-LRA, DPI, and DCR-Diff all assume No-box-calibrated threat models. This is line with many results from the MIA literature where \cite{watson2022on} showed that reference datasets are theoretically important for calibrating attacks and \cite{vanbreugel2023membership} showed improvements in DOMIAS by considering the calibrated threat model. 

\subsubsection{Measures of Privacy Leakage}
A widely used metric for privacy and overfitting evaluation in tabular generative models is Distance to Closest Record- Proportion \footnote{Note, this is often called DCR in the synthetic data literature. However, \cite{ganleaks} proposes an MIA where the scoring function is a distance computation for a test point and a synthetic point that has been called DCR in MIA papers. To avoid confusion, we refer to the similarity metric DCR as DCR-Prop and the MIA as DCR.} \citep{park2018data, lu2019empirical, yale2019assessing, zhao2021ctab,guillaudeux2023patient, liu2023tabular,solatorio2023realtabformer, tabsyn}, which compares the distance from each training point to its nearest neighbor in the synthetic and reference datasets. A proportion summary statistic is then computed with the intuition being that if the synthetic data are usually closer to the training versus the reference set this implies privacy leakage. For example, a DCR-Prop score of 0.50 would suggest that the synthetic data is evenly-distributed amongst the training and reference observations, suggesting a private dataset.
\begin{figure}
    \centering
    \includegraphics[width=1\linewidth]{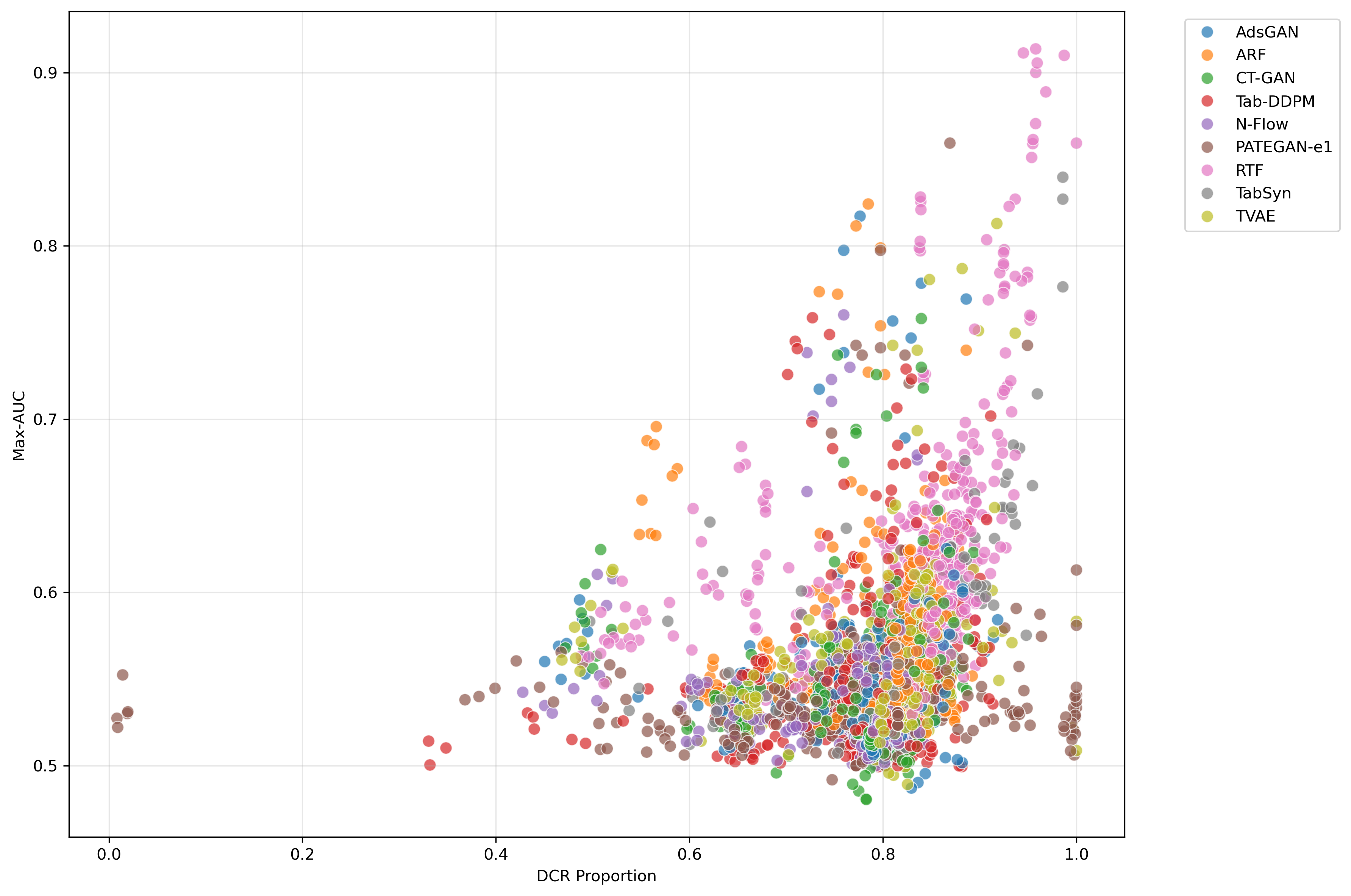}
    \caption{A plot of the Max-AUC of all MIAs and the DCR-Proportion for each synthetic dataset. We find that these two privacy metrics are weakly correlated ($r=.225,  p=5.9e-13$) suggesting that DCR-Proportion is an inadequate measure of privacy and overfiting.}
    \label{fig:dcr_privacy}
\end{figure}
We find that architectures that exhibit the most MIA privacy leakage have the best DCR-prop scores (See Table \ref{tab:metrics_privacy}). We evaluate the correlation of the Max-AUC for all synthetic datasets from each model and seed with the corresponding DCR-Prop score and find that DCR-Prop \textbf{is weakly correlated with the Max-AUC} (see Figure \ref{fig:dcr_privacy}). This shows that Distance to Closest Record Proportion which is the most common privacy metric used in the tabular synthesis community is basically uncorrelated with MIA privacy risk ($r=.225,  p=5.9e-13$).

\begin{table*}[h]
\centering
\caption{Mean ranks for each MIA across experiment unit with completion times in seconds. Gen-LRA and DPI outperform other attacks, but the high standard deviations indicate that there is considerable variability in performance. Values are shown as Mean (Standard Deviation).}
\begin{tabular}{lcccccc}
\toprule
\textbf{MIA} & \textbf{AUC-ROC} & \textbf{TPR@FPR=0} & \textbf{TPR@FPR=0.001} & \textbf{TPR@FPR=0.01} & \textbf{TPR@FPR=0.1} & \textbf{Run Time}\\
\midrule
Classifier              & 4.92 (2.77) & 6.19 (2.49) & 5.82 (2.59) & 5.42 (2.62) & 5.35 (2.71) & 2.77 (5.04)\\
DCR                    & 4.93 (2.52) & 5.03 (2.65) & 5.08 (2.79) & 4.97 (2.71) & 5.03 (2.59) & 0.14 (0.29)\\
DCR-Diff               & 4.57 (2.66) & \underline{4.20 (2.37)} & \underline{4.53 (2.71)} & 4.53 (2.84) & 4.60 (2.82) & 0.15 (0.31)\\
DOMIAS                 & 5.23 (2.45) & 4.59 (2.30) & 4.87 (2.65) & 5.46 (2.91) & 5.51 (2.84) & 0.53 (1.43)\\
DPI                    & \textbf{3.60 (2.31)} & 6.52 (2.49) & 5.39 (2.46) & \underline{4.38 (2.27)} & \underline{3.85 (2.27)} & 0.12 (0.27)\\
Density Estimate   & 5.84 (2.68) & 4.65 (2.42) & 4.98 (2.71) & 5.33 (2.72) & 5.51 (2.65) & 0.61 (1.571)\\
Gen-LRA                 & \underline{4.00 (2.63)} & \textbf{3.72 (2.71)} & \textbf{3.64 (2.56)} & \textbf{3.81 (2.61)} & \textbf{3.73 (2.60)} & \textbf{27.14 (68.67)}\\
LOGAN                  & 6.40 (3.23) & 5.11 (2.56) & 5.60 (2.95) & 6.11 (2.97) & 6.21 (2.98) & 1.51 (0.67)\\
Local Neighborhood     & 6.47 (2.20) & 6.20 (2.34) & 6.10 (2.45) & 6.05 (2.43) & 6.11 (2.25) & 0.05 (0.16)\\
MC                     & 5.96 (2.42) & 5.37 (2.32) & 5.63 (2.62) & 5.82 (2.54) & 5.96 (2.61) & 1.19 (4.29)\\
\bottomrule
\end{tabular}

\label{tab:mia_performance}
\end{table*}

\subsubsection{Differential Privacy as Defense}

Finally, we examine how Differential Privacy (DP)~\cite{dwork2006calibrating} affects privacy preservation against these attacks. We initialize PATEGAN with privacy budgets of $\epsilon = \{1,5,10\}$. Our results show that differential privacy does provide aggregate defense against MIAs compared to high-fidelity models without DP protections (Figure~\ref{fig:privacy_utility}). However, to assess the \textit{reliability} of this protection, we examine the consistency of PATEGAN by reporting mean Max MIA performance across the 10 datasets where attacks were most successful (Table~\ref{tab:pategan_privacy}). While models with $\epsilon=1$ demonstrate superior privacy protection compared to higher $\epsilon$ values, they remain vulnerable to privacy breaches in worst-case scenarios. For instance, if hospital patients could be identified with a 19\% true positive rate and zero false positives, practitioners would reasonably conclude the dataset lacks adequate privacy protection. This finding demonstrates that even conservative differentially private implementations can exhibit privacy leakage, highlighting how Synth-MIA enables practitioners to identify and mitigate such risks in real-world deployments.

\section{Discussion}
\subsection{Synth-MIA as an Interpretability Method}

MIAs give interpretable information about model failure as the success of the attack is derived from a failure in the model in that it is memorizing training examples or overfitting (see Table \ref{tab:mia_summary}). The results of this experiment imply that models with greater synthetic data quality are accomplishing this by implicitly inducing training failure (see Figure \ref{fig:privacy_utility}). If producing better quality synthetic data comes with a privacy cost, this needs to be emphasized in tabular model research. Synth-MIA can thus help lead to the development of better models, as if an architecture can provide higher quality synthetic data without greater training failure this and thus privacy leakage, it would be a distinct advantage researchers and practitioners should prefer. 

Another finding from these results is that there may not be a "one size fits all" tabular generative model for all use-cases and datasets. Table \ref{tab:metrics_privacy} shows that even privacy-orientated generators like Ads-GAN and PATEGAN can leak training data at very low fixed-FPR levels. While these TPR@FPR values are much lower than in the supervised learning setting, depending on the domain of successfully classifying the positive membership of ~4\% of a test set without a false positive could qualify as unacceptable privacy risk. Figure \ref{fig:training_size} also shows that the dataset itself matters when it comes to privacy and that practitioners should endeavor to use as much data as they can in training as it implies greater privacy protection. Indeed, there may be a niche for researchers to develop tabular models for different use-case and dataset settings. 

Lastly, Synth-MIA can enable further experimental designs to evaluate how and why different data generators may leak privacy and in what situations. While our benchmarking follows a correlative study design, Synth-MIA can be used by tabular generative model researchers as a mechanistic Interpretability tool to study causal factors in model failure.
\subsection{Non-Adversarial Measures of Privacy}

DCR-Prop is an inadequate metric for assessing privacy risks in synthetic data release, as it assumes no adversary, lacks a clearly defined threat model, and does not correspond to any material privacy risk \cite{ganev2023inadequacy, ward2024dataplagiarismindexcharacterizing,Gen-LRA}. As demonstrated in Table \ref{tab:metrics_privacy}, TabSyn achieves the same mean DCR-Proportion score as TVAE, even though it often has a rougly 25\% higher TPR@FPR values. Use of DCR-Proportion can lead to an incorrect understanding of the privacy properties of a current State-of-the-Art model. Indeed, Figure \ref{fig:dcr_privacy} presents the results of the Pearson correlation between the Max-AUC across the synthetic datasets from all models, datasets, and seeds to the corresponding DCR-Proportion, finding that these two measures are only weakly correlated. 

While DCR-Prop has been used to assess model overfitting, it fails to characterize the consequences of overfitting in any meaningful way. If a practitioner generates high-quality synthetic data with a high DCR-Prop, there is no clear reason why this should be a concern. In contrast, MIAs not only can quantify model overfitting but also establish its connection to real-world privacy risks, providing a more actionable measure for responsible data practitioners. We therefore recommend replacing DCR-Prop with MIA-based privacy auditing to ensure a more rigorous and adversarially aware evaluation of synthetic data generators.

\begin{table}
\centering
\caption{Mean max MIA performance metrics over the 10 datasets with the highest privacy leakage for various $\epsilon$ levels of PATEGAN. While PATEGAN protects privacy broadly aggregated over many datasets and experiment runs, there are still individual instances where a differentially private synthetic data generator can fail to effectively defend against MIAs.}

\begin{tabular}{lccccc}
\toprule
\cmidrule(lr){2-4}
 \textbf{Model} & \textbf{AUC-ROC} & \textbf{TPR@FPR=0} &  \textbf{TPR@FPR=0.1} \\
\midrule
{$\epsilon=1$}       & 0.628 (0.079) & 0.190 (0.174) &  0.298 (0.130) \\
{$\epsilon=5$}   & 0.670 (0.074) & 0.248 (0.165) &  0.328 (0.120) \\
{$\epsilon=10$}  & 0.668 (0.079) & 0.254 (0.166) &  0.322 (0.129) \\
\bottomrule
\end{tabular}
\label{tab:pategan_privacy}
\end{table}
\subsection{Role of Synth-MIA in Synthetic Data Deployment}
Privacy auditing of synthetic data using MIAs cannot be effectively conducted using a single attack, as no strictly dominant attack exists. Since an adversary cannot predict which attack will be most effective, practitioners responsible for data release should adopt a conservative approach by evaluating many attack strategies. Different attacks may exploit distinct weaknesses in the synthetic data generation process, and no single method is universally optimal. By systematically applying a range of MIAs using a testbed, practitioners can make more informed decisions regarding the balance between privacy protection and data utility before releasing synthetic datasets. Indeed, in differentially private pipelines, Synth-MIA can serve as a final check to empirically evaluate the performance of synthetic data.

Synth-MIA supports a dynamic, feedback-driven approach to privacy preservation in synthetic data generation by continuously evaluating how model modifications impact privacy leakage, enabling organizations to iteratively improve their models while maintaining alignment with evolving privacy standards. It provides a rigorous framework for quantifying privacy risks through a comprehensive suite of membership inference attacks and privacy metrics, offering clear numerical indicators that facilitate informed decision-making. This transparency is essential for stakeholders, as it clarifies the trade-offs between data utility and privacy, thereby strengthening privacy assurance protocols. Additionally, Synth-MIA bridges the gap between technical teams and regulatory or legal departments by generating standardized, reproducible privacy audit reports that are accessible and meaningful to both groups. Privacy Machine Learning Engineers can leverage detailed metrics to fine-tune generative models, while Legal and Customer Trust teams gain concrete evidence that privacy risks are systematically monitored and mitigated, fostering trust and facilitating smoother internal collaboration. By embedding an iterative process of privacy enhancement, Synth-MIA can not only improve the quality and safety of synthetic data deployments but also promotes a proactive privacy management culture within organizations, ensuring that innovation and compliance progress in tandem.

\subsection{Limitations}
Synth-MIA has several limitations that warrant further investigation. The auditing procedure, while designed to capture diverse failure modes, can be computationally demanding depending on the choice of attacks and the size of the evaluation dataset. In our experiments across 57 datasets, the average runtime per iteration was approximately 3 minutes, with a maximum of 10 minutes on the largest dataset when executed on an AWS g5-xlarge EC2 instance. However, the majority of this computational cost was attributed to Gen-LRA. To accommodate resource-constrained environments, Synth-MIA allows users to selectively enable specific attacks, providing flexibility in balancing performance and efficiency.

Another key limitation of Synth-MIA is its reliance on generator-agnostic attacks, which restricts its applicability to certain adversarial threat models. Specifically, it does not incorporate white-box or model-informed shadow-box attacks. While this constraint limits compatibility with certain attack paradigms, we argue that No-box and No-box-calibrated attacks are most representative of real-world adversaries targeting publicly released synthetic datasets. Moreover, by emphasizing these attack paradigms, Synth-MIA remains simpler and thus more accessible to a broader audience, including those without specialized expertise in adversarial machine learning, thereby fostering its adoption within the synthetic data community.

Lastly, a final limitation of privacy auditing with MIAs is that there may exist a more powerful attack that has not been discovered and thus current methods may underestimate the actual achievable privacy risk by a future adversary. While a real risk, this is not unique to Synth-MIA and this software can be used by researchers to develop and share new attacks that can facilitate tighter privacy auditing in the future. This uncertainty about unknown future attacks is a compelling reason why practitioners should prefer differentially private synthetic data generation methods which provide formal privacy guarantees that hold regardless of what attacks may be developed in the future when correctly implemented.

\section{Conclusion}
Synth-MIA is a robust and flexible framework and testbed designed to facilitate principled privacy auditing for tabular synthetic data under a realistic threat model. r providing a comprehensive suite of tools, Synth-MIA supports benchmarking, the development of new attack methodologies, and in-depth privacy assessments. Through the largest synthetic data privacy experiment conducted to date, we demonstrate that Synth-MIA is a powerful tool for studying the behavior of tabular generative models and evaluating existing privacy metrics. Here, we find that Synth-MIA better characterizes a quality versus privacy trade-off than other methods. While Synth-MIA is not perfect in that it does not guarantee finding the theoretical maximum vulnerability of a model and the computational cost of Synth-MIA varies depending on the attack strategies employed, Synth-MIA is the most effective solution currently available and its customizable design allows users to adapt the framework to their specific computational constraints. Future research should focus on improving attack methodologies and developing approaches that can approximate or determine the theoretical maximum privacy risk. Synth-MIA serves as a foundational tool for advancing this line of research, enabling a deeper understanding of privacy vulnerabilities in synthetic data and guiding the development of stronger privacy-preserving techniques.

\bibliographystyle{ACM-Reference-Format}
\bibliography{main}

\newpage
\appendix

\onecolumn

\section{Datasets}
We report the data sets used for the experiments in Sections 5 in Table \ref{tab:datasets}.
\begin{table}[H]
    \caption{List of Datasets included in Section 5.}
    \centering
    \small
\begin{tabular}{lrrrrr}
\hline
Dataset & OpenML ID & N-size & Classes &  Cat. Feat. &  Num Feat. \\
\hline
GesturePhaseSegmentationProcessed & 4538 & 9873 & 5 & 1 & 32 \\
MiceProtein & 40966 & 1080 & 8 & 5 & 77 \\
PhishingWebsites & 4534 & 11055 & 2 & 31 & 0 \\
adult & 1590 & 48842 & 2 & 9 & 6 \\
analcatdata\_authorship & 40983 & 4839 & 2 & 1 & 5 \\
analcatdata\_dmft & 469 & 797 & 6 & 5 & 0 \\
bank-marketing & 1461 & 45211 & 2 & 10 & 7 \\
banknote-authentication & 1462 & 1372 & 2 & 1 & 4 \\
blood-transfusion-service-center & 1464 & 748 & 2 & 1 & 4 \\
breast-w & 15 & 699 & 2 & 1 & 9 \\
car & 40975 & 1728 & 4 & 7 & 0 \\
churn & 40701 & 5000 & 2 & 5 & 16 \\
climate-model-simulation-crashes & 1467 & 540 & 2 & 1 & 20 \\
cmc & 23 & 1473 & 3 & 8 & 2 \\
connect-4 & 40668 & 67557 & 3 & 43 & 0 \\
credit-approval & 29 & 690 & 2 & 10 & 6 \\
credit-g & 31 & 1000 & 2 & 14 & 7 \\
cylinder-bands & 6332 & 540 & 2 & 22 & 18 \\
diabetes & 37 & 768 & 2 & 1 & 8 \\
dresses-sales & 23381 & 500 & 2 & 12 & 1 \\
electricity & 151 & 45312 & 2 & 2 & 7 \\
eucalyptus & 43924 & 736 & 5 & 15 & 5 \\
first-order-theorem-proving & 1475 & 6118 & 6 & 1 & 51 \\
ilpd & 1480 & 583 & 2 & 2 & 9 \\
jm1 & 1053 & 10885 & 2 & 1 & 21 \\
kc1 & 1067 & 2109 & 2 & 1 & 21 \\
kc2 & 1063 & 522 & 2 & 1 & 21 \\
kr-vs-kp & 3 & 3196 & 2 & 37 & 0 \\
letter & 6 & 20000 & 26 & 1 & 16 \\
mfeat-fourier & 14 & 2000 & 10 & 1 & 76 \\
mfeat-karhunen & 16 & 2000 & 10 & 1 & 64 \\
mfeat-morphological & 18 & 2000 & 10 & 1 & 6 \\
mfeat-zernike & 22 & 2000 & 10 & 1 & 47 \\
numerai28.6 & 23517 & 96320 & 2 & 1 & 21 \\
optdigits & 28 & 5620 & 10 & 1 & 64 \\
ozone-level-8hr & 1487 & 2534 & 2 & 1 & 72 \\
pc3 & 1044 & 10936 & 3 & 4 & 24 \\
pendigits & 32 & 10992 & 10 & 1 & 16 \\
phoneme & 1489 & 5404 & 2 & 1 & 5 \\
qsar-biodeg & 1494 & 1055 & 2 & 1 & 41 \\
satimage & 182 & 6430 & 6 & 1 & 36 \\
segment & 40984 & 2310 & 7 & 1 & 19 \\
sick & 38 & 3772 & 2 & 23 & 7 \\
spambase & 44 & 4601 & 2 & 1 & 57 \\
splice & 46 & 3190 & 3 & 62 & 0 \\
steel-plates-fault & 40983 & 4839 & 2 & 1 & 5 \\
texture & 40499 & 5500 & 11 & 1 & 40 \\
tic-tac-toe & 50 & 958 & 2 & 10 & 0 \\
vehicle & 54 & 846 & 4 & 1 & 18 \\
\hline
\end{tabular}
    \label{tab:datasets}
\end{table}

\newpage
\section{Experiment Hyperparameters}
\subsection{Model Configuration Details}

All tabular generative models were trained and evaluated using their default hyperparameters for both training and sampling procedures. The models CT-GAN, TVAE, N-Flows, ARF, Tab-DDPM, AdsGAN, and PATEGAN were implemented using the Synthcity framework (\url{https://synthcity.readthedocs.io/en/latest/}). The state-of-the-art models TabSyn and REaLTabFormer (RTF) were implemented using their native repositories (\url{https://github.com/amazon-science/tabsyn} and \url{https://github.com/worldbank/REaLTabFormer}, respectively).

The only modification to default hyperparameters was made for PATEGAN, where we varied the differential privacy parameter $\epsilon \in \{1, 5, 10\}$ to evaluate privacy-utility trade-offs while maintaining all other default settings.
\subsection{Attack Configuration Details}

Table~\ref{tab:attack_hyperparams} summarizes the hyperparameter settings for all membership inference attacks evaluated in our framework.

\begin{table}[h!]
\centering
\caption{Hyperparameter configurations for all membership inference attacks. KDE = Kernel Density Estimation.}
\begin{tabular}{llp{8cm}}
\toprule
\textbf{Attack} & \textbf{Distance Metric} & \textbf{Key Hyperparameters} \\
\midrule
\textsc{Gen-LRA} & --- & $k \in \{1, 5, 10, 25, 50, 100\}$; likelihood computed via kernel density estimation (KDE) \\
\textsc{DPI} & L2 & $k \in \{1, 5, 10, 25, 50, 100\}$; nearest neighbors identified using L2 distance \\
\textsc{DCR}, \textsc{DCR-Diff} & L2 & -- \\
\textsc{Local Neighborhood} & L2 & Radius $= 1.0$ \\
\textsc{Classifier} & --- & Random forest with scikit-learn default parameters \\
\textsc{LOGAN} & --- & Hidden\_dim = 256, learning\_rate = 1e-3, batch\_size = 256 \\
\textsc{MC} & L2 & -- \\
\textsc{Density Estimator} & --- & Density estimated via KDE \\
\textsc{DOMIAS} & --- & Density ratios estimated via KDE \\
\bottomrule
\end{tabular}

\label{tab:attack_hyperparams}
\end{table}

\noindent Our use of KDE follows the findings of \cite{Gen-LRA}, which demonstrated its superior performance compared to deep neural network-based density estimation in membership inference settings.

\section{Auditing Privacy in Tabular Generative Models}

\textbf{Definition 1 (Auditing Setup for Tabular Generative Models).}  
Let $G$ be a tabular generative model trained on a private dataset $D_{\text{train}}$, generating a synthetic dataset $S = G(D_{\text{train}})$. Let $R$ be a reference dataset that follows the same distribution as $D_{\text{train}}$ but does not contain any records from it. The \textbf{privacy auditing procedure} consists of two components:

\begin{enumerate}
    \item \textbf{Game} $\mathcal{A}: \mathcal{X} \to \{0,1\}$ – An adversary observes a record $x$ from $S$ (\textbf{black-box MIA}) or from both $S$ and $R$ (\textbf{shadow-box MIA}) and attempts to infer whether $x$ was part of $D_{\text{train}}$.
    \item \textbf{Evaluation} $\Phi: \mathcal{O} \times \mathcal{F} \to \{0,1\}$ – Given an observation from $\mathcal{A}$, the audit mechanism tests whether the generative model violates a predefined \textbf{privacy hypothesis} $f \in \mathcal{F}$, where $\mathcal{F}$ is a family of privacy constraints defining acceptable membership inference risks.
\end{enumerate}

\textbf{Toy Example of a Privacy Hypothesis.}  
A simple privacy hypothesis $f_{\text{mem}}$ states that a generative model should not produce records that are too similar to training samples. We define:
\begin{itemize}
    \item If a synthetic record $x$ has a nearest neighbor in $D_{\text{train}}$ within a Euclidean distance of \textbf{0.01}, we reject the privacy hypothesis (i.e., the model memorizes exact samples).
    \item Otherwise, we accept the privacy hypothesis, considering the model sufficiently privacy-preserving.
\end{itemize}

The audit procedure is said to be \textbf{$\psi$-accurate} if, for any generative model satisfying a given privacy hypothesis $f$, the probability of correctly accepting the hypothesis is at least $\psi$:

$$
\Pr_{x \sim \mathcal{A}(G(D_{\text{train}}))} [\Phi(x, f) = 1] \geq \psi.
$$

\textbf{Definition 2 (Membership Inference Attack for Generative Models).}  
A \textbf{membership inference attack (MIA)} against a tabular generative model is a statistical test evaluating whether a given sample $x$ was present in $D_{\text{train}}$. The attack framework consists of:
\begin{itemize}
    \item \textbf{Black-box MIA}: The adversary uses only $S$ to infer membership.
    \item \textbf{Shadow-box MIA}: The adversary utilizes both $S$ and $R$ to improve inference performance.
\end{itemize}

The adversary constructs a decision function $\Lambda: \mathcal{X} \to \mathbb{R}$ based on a likelihood ratio test:

$$
\Lambda(x) = \frac{\Pr(x \mid G(D_{\text{train}}))}{\Pr(x \mid R)}
$$

Membership is predicted if $\Lambda(x) \geq \tau$, where $\tau$ is a decision threshold chosen based on statistical calibration (e.g., maximizing attack precision while maintaining a low false positive rate).

The success of the attack is quantified by the \textbf{true positive rate (TPR)} and \textbf{false positive rate (FPR)}:

$$
\text{TPR} = \Pr[\Lambda(x) \geq \tau \mid x \in D_{\text{train}}],
$$

$$
\text{FPR} = \Pr[\Lambda(x) \geq \tau \mid x \in R].
$$

A \textbf{generative model is considered privacy-risky} if MIA performance significantly exceeds a baseline threshold determined by privacy-preserving models.

\textbf{Definition 3 (Privacy Hypothesis and Empirical Privacy Assessment).}  
The \textbf{privacy hypothesis} $f$ formalizes the expected robustness of a generative model against membership inference attacks. A generative model $G$ is said to satisfy privacy hypothesis $f$ if the observed membership inference risk remains below a specified bound. We define $f$ in terms of the trade-off between TPR and FPR:

$$
f(\text{FPR}) = \sup_{\tau} \text{TPR}(\tau).
$$

A common choice of $f$ is an \textbf{empirical differential privacy bound}, such as:

$$
\text{TPR}(\tau) \leq e^\epsilon \cdot \text{FPR}(\tau) + \delta.
$$

\textbf{Definition of $\tau$.}  
Here, $\tau$ is a calibrated threshold that determines when the adversary predicts membership. It can be selected using statistical criteria, such as:
\begin{itemize}
    \item \textbf{Maximizing precision}: $\tau$ is chosen to balance the trade-off between TPR and FPR.
    \item \textbf{Fixed quantiles}: $\tau$ is set at a specific quantile of the observed likelihood ratio distribution.
    \item \textbf{Cross-validation}: $\tau$ is chosen based on validation data to minimize false positives.
\end{itemize}

\textbf{Empirical Privacy Estimation.}  
The \textbf{empirical privacy} of $G$ is assessed by selecting the \textbf{strongest privacy hypothesis} that is not rejected by the audit:

$$
f^* = \max \{ f \in \mathcal{F} \mid \Phi(\mathcal{A}(G(D_{\text{train}})), f) = 1 \}.
$$

If $f^*$ falls below a pre-specified threshold (e.g., corresponding to acceptable differential privacy guarantees), the generative model is flagged as \textbf{having training data privacy concerns}. This approach systematically quantifies privacy leakage risks and provides a framework for evaluating different tabular generative models.

\subsection{Justification for Using Multiple MIAs in Empirical Differential Privacy Auditing}

\textbf{Empirical Differential Privacy as a Supremum Over Attacks.}  
One of the key aspects of auditing privacy in generative models is assessing the worst-case membership inference risk. A common approach in empirical differential privacy (EDP) is to evaluate the maximum attack success rate across a broad set of membership inference attacks (MIAs). This methodology is based on the theoretical notion that the true empirical privacy level of a model is determined by the supremum over all possible attacks. That is, if an adversary were to employ the most effective attack strategy, it would yield the tightest possible empirical privacy bound. Therefore, taking the max AUC or max TPR@FPR across a diverse set of MIAs provides a robust approximation of this supremum in practice.

\textbf{Mathematical Justification for Maximizing Attack Performance.}  
In a formal privacy auditing framework, the empirical privacy guarantee of a generative model should not be dictated by a single attack but by the most effective attack within a large class of MIAs. Formally, let $\mathcal{A} = \{A_1, A_2, \dots, A_n\}$ represent a set of different membership inference attacks. The empirical differential privacy level of the generative model should be measured as:

$$
\sup_{A \in \mathcal{A}} \text{TPR}_A(\tau) \quad \text{subject to } \text{FPR}_A(\tau) \leq \delta.
$$

This means that the worst-case privacy leakage is estimated by selecting the most effective MIA, which naturally leads to reporting the maximum observed attack performance across different strategies. Since different MIAs exploit different aspects of overfitting—such as direct memorization, attribute leakage, and distributional biases—this approach ensures that privacy assessments are comprehensive and robust.

\textbf{Practical Benefits of Running Multiple MIAs.}  
From a practical perspective, running multiple MIAs is crucial for obtaining a tight empirical privacy estimate. First, different MIAs capture different types of privacy vulnerabilities; some attacks are sensitive to exact data memorization, while others focus on distributional similarities between the synthetic and training data. Second, relying on a single attack may lead to an underestimation of privacy risk, as no single method is universally optimal. Lastly, the principle of empirical differential privacy is fundamentally adversarial—in real-world applications, attackers seek the most effective strategies, and a robust audit should reflect this worst-case scenario. By taking the maximum attack success rate across multiple MIAs, we obtain a more precise, adversary-aware empirical privacy bound, ensuring that the audit reflects realistic and meaningful privacy risks.
\section{Additional Figures}
\begin{figure}[H]
    \centering
    \includegraphics[width=1\linewidth]{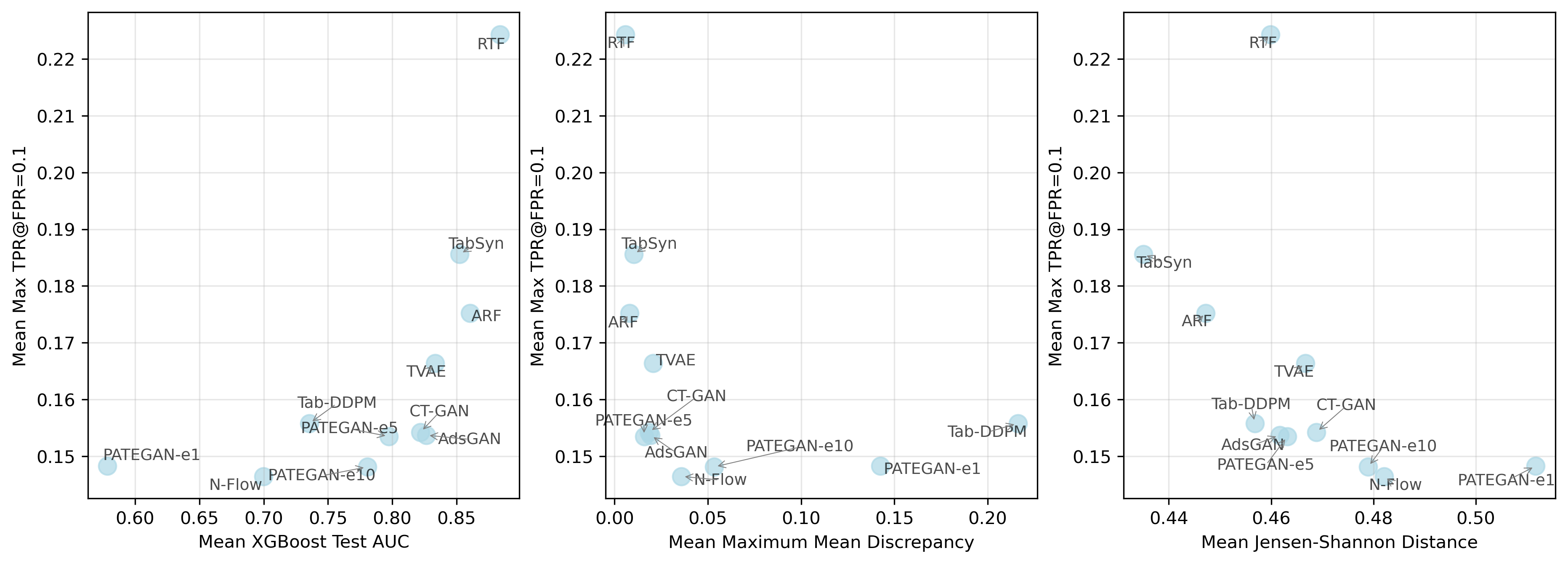}
    \caption{Privacy and quality trade-off across tabular generative model architectures. We plot the mean Max-TPR@FPR=0.1 of all attacks for all datasets for each model architecture against corresponding measures of mean Classifier AUC, mean Maximum Mean Discrepancy, and mean Jensen-Shannon Distance.}
\end{figure}
\begin{figure}[H]
    \centering
    \includegraphics[width=1\linewidth]{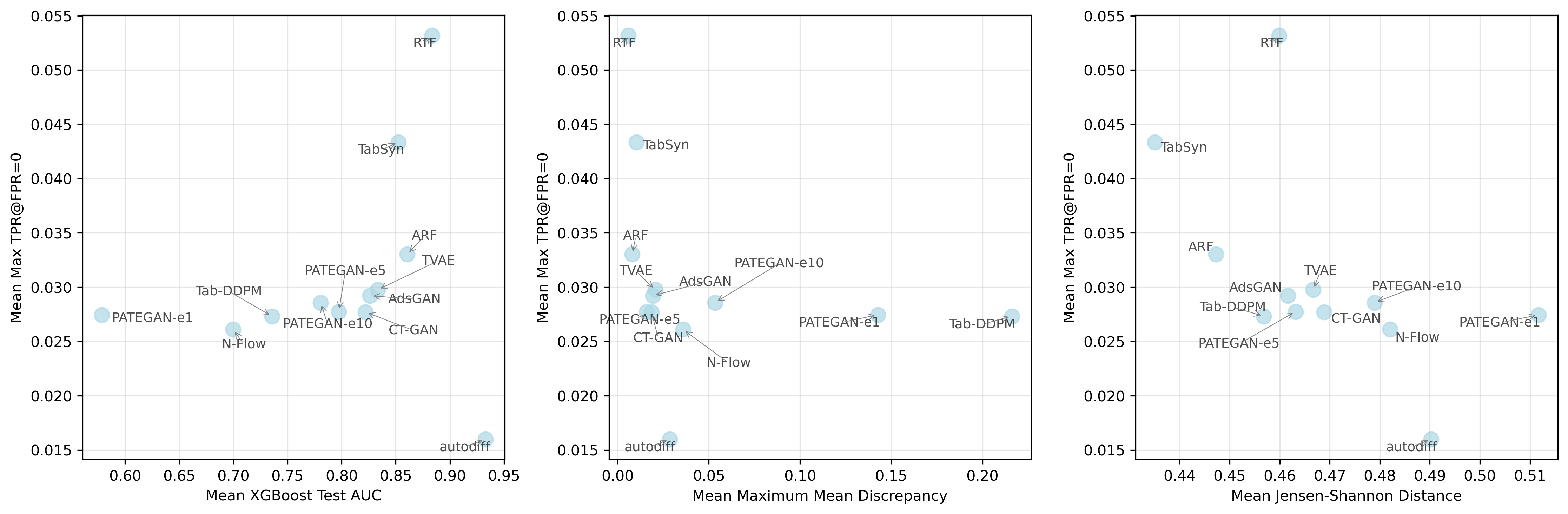}
    \caption{Privacy and quality trade-off across tabular generative model architectures. We plot the mean Max-TPR@FPR=0 of all attacks for all datasets for each model architecture against corresponding measures of mean Classifier AUC, mean Maximum Mean Discrepancy, and mean Jensen-Shannon Distance.}
    \end{figure}
\begin{figure}[H]
    \centering
    \includegraphics[width=0.4\linewidth]{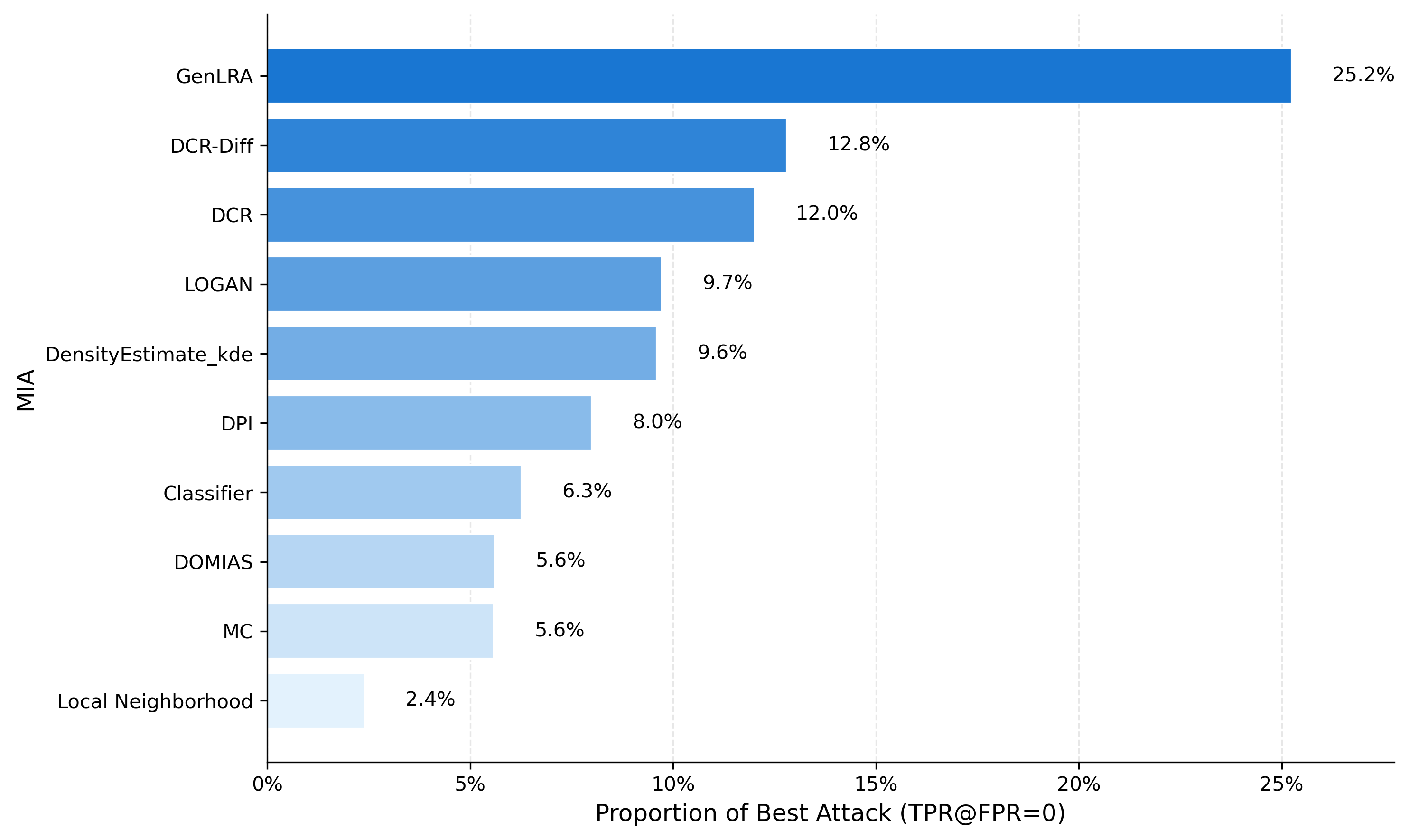}
    \caption{Proportion of best attack across all generative models, datasets for True Positive Rate at False Positive Rate = 0. }
\end{figure}

\begin{figure}[H]
    \centering
    \includegraphics[width=0.4\linewidth]{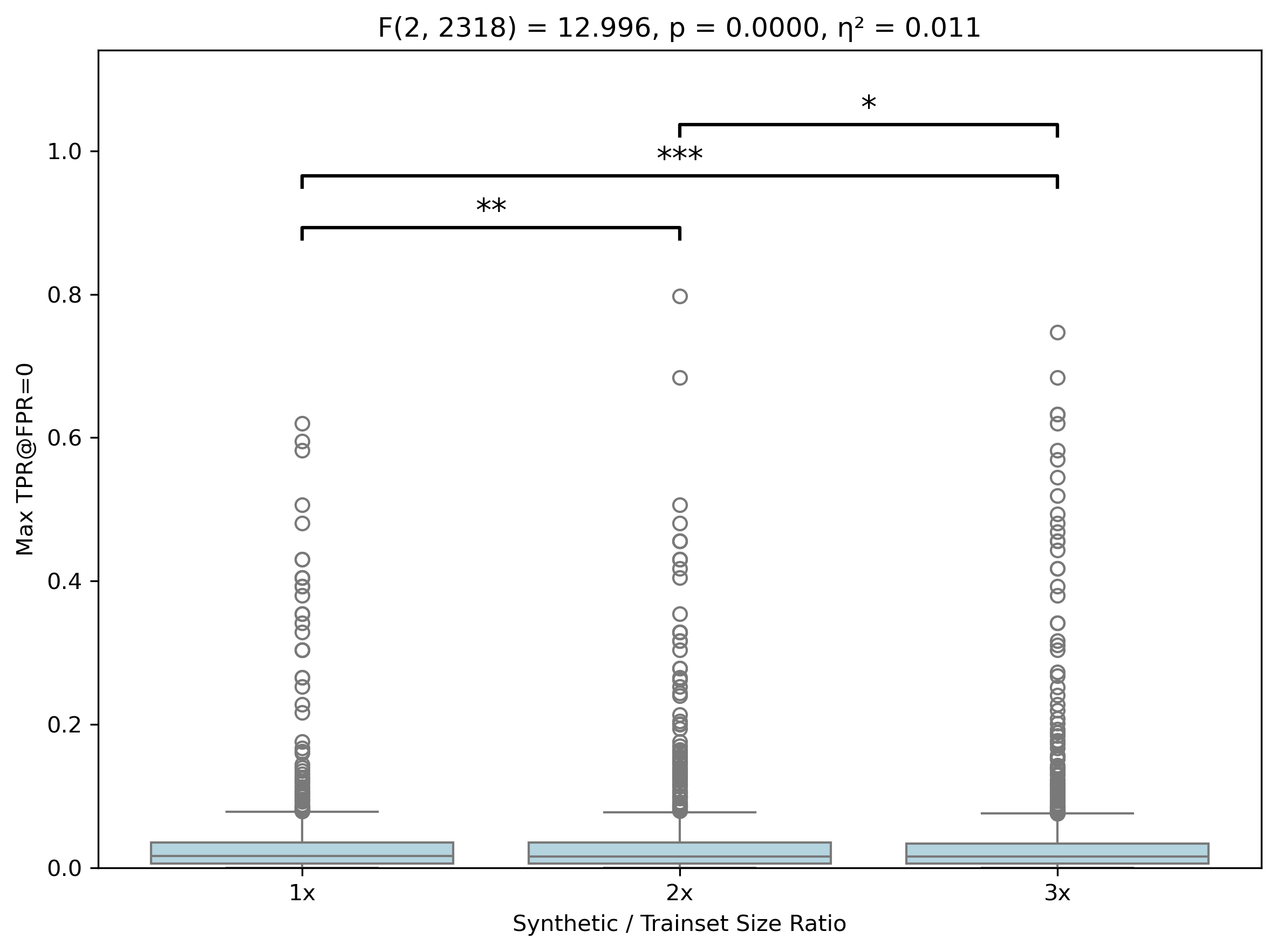}
    \caption{One-Way Repeated Measures ANOVA of Max-TPR@FPR=0 for various synthetic data sizes. We find that there are statistically significant differences in the synthetic dataset sizes but the effect size is small.}
\end{figure}

\begin{figure}[H]
    \centering
    \includegraphics[width=0.4\linewidth]{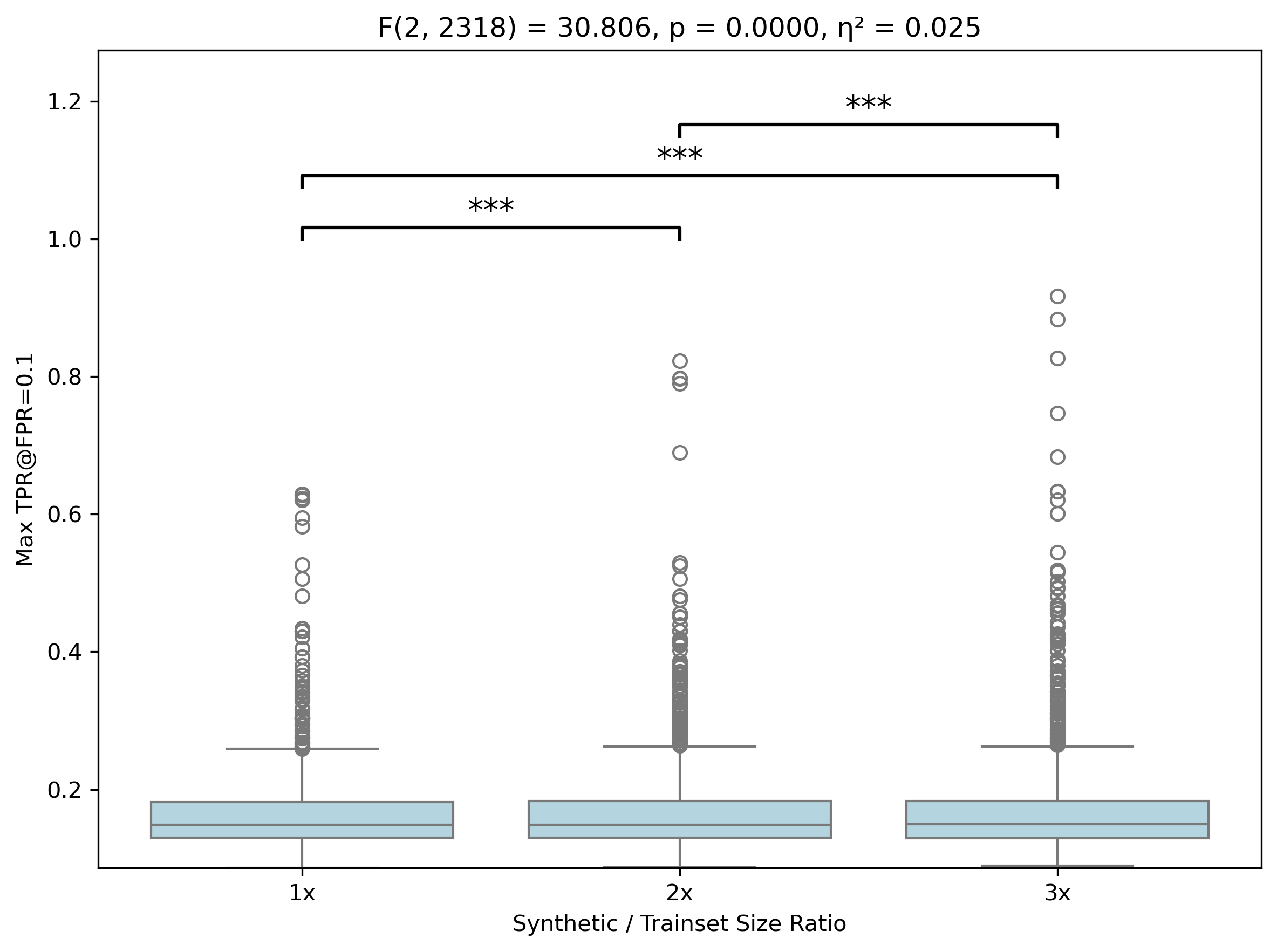}
    \caption{One-Way Repeated Measures ANOVA of Max-TPR@FPR=0.1 for various synthetic data sizes. We find that there are statistically significant differences in the synthetic dataset sizes but the effect size is small.}
\end{figure}


\end{document}